\documentclass[english, hidelinks]{article}
\usepackage[T1]{fontenc}
\usepackage[latin9]{inputenc}
\usepackage{amsmath}
\usepackage{amsthm}
\usepackage{amssymb}
\usepackage{graphicx}
\usepackage{hyperref}
\usepackage[ruled]{algorithm}
\usepackage{xcolor}

\makeatletter
\theoremstyle{plain}

\usepackage{babel}
\providecommand{\theoremname}{Theorem}

\usepackage{babel}
\providecommand{\theoremname}{Theorem}

\makeatother

\usepackage{babel}
\providecommand{\theoremname}{Theorem}

\begin{document}
\title{Quasi-Objective Eddy Visualization from Sparse Drifter Data}
\author{Alex P. Encinas Bartos, Nikolas O. Aksamit
and George Haller\thanks{Email:georgehaller@ethz.ch}\\
 Institute for Mechanical Systems\\
 ETH Zürich\\
 8092, Zürich, Switzerland\\
}
\maketitle
\begin{abstract}
We employ a recently developed single-trajectory Lagrangian diagnostic tool, the trajectory rotation average $ (\mathrm{\overline{TRA}}) $, to visualize oceanic vortices (or eddies) from sparse drifter data. 
We apply the $ \mathrm{\overline{TRA}} $ to two drifter data sets that cover various oceanographic scales: the Grand Lagrangian Deployment (GLAD) and the Global Drifter Program (GDP). Based on the $ \mathrm{\overline{TRA}} $, we develop a general algorithm that extracts approximate eddy boundaries. We find that the $ \mathrm{\overline{TRA}} $ outperforms other available single-trajectory-based eddy detection methodologies on sparse drifter data and identifies eddies on scales that are unresolved by satellite-altimetry.
\end{abstract}
%


\begin{quote}
\textbf{~~~~~~~~~~~~~~~~~~~~~~~~~~~~~~~~~~~~~~~~~~~~~~~~~~~~~Summary}
\\ \\
\textbf{Meso and submesoscales vortices (or eddies) can trap and transport material over large distances, thereby playing a crucial role in the dynamics of our ecosystem. In order to expand our understanding of the transport of marine tracers, we need to accurately and reliably track the evolution of vortical flow structures. Drifter trajectories represent a valuable but sparse source of information for this purpose. We utilize this information here to evaluate a single-trajectory Lagrangian diagnostic tool that approximates the local material rotation in the flow in a quasi-objective fashion. Our findings on two distinct data sets suggest that this diagnostic accurately highlights material vortices from sparse drifter data.}
\end{quote}

\section{Introduction}

Oceanic vortices (or eddies) are highly energetic coherent flow structures that can transport material over large distances.
Studying the dynamics of eddies is key to understanding the dispersion of marine tracers, such as biological nutrients and pollutants \cite{VanSebille2012, Froyland2014, Beron2015, Miron2020, Abernathey2021}. Lagrangian eddies, generally referred to as elliptic Lagrangian coherent structures (LCS) in the dynamical systems literature \cite{Haller2015}, are material objects that trap and transport floating particles over large distances in the ocean in a coherent fashion. The size of such eddies ranges from a few kilometers (submesoscale) to hundreds of kilometers (mesoscale). \\ \\
Mesoscale eddies have predominantly been inferred from the sea-surface-height (SSH) field derived from satellite-altimetry data \cite{Chelton2007, Faghmous2015, Abernathey2018}. There is, however, increasing evidence that submesoscale currents on the order of a few kilometers influence the marine ecosystem at least as strongly as mesoscale eddies, especially along coastlines and oceanic fronts \cite{McWilliams2016, Levy2018}. Despite this observation, submesoscale eddies are rarely studied in detail because their footprint in the SSH field is weak or nonexistent. In contrast to the SSH field, however, drifters follow ocean currents  closely and hence resolve small scale features accurately \cite{Lumpkin2016}. In addition, surface drifters provide information about the local ocean surface velocity field at a very high temporal resolution. Ocean drifters are, however, sparsely distributed in space, rendering most LCS diagnostics inapplicable to their trajectories. Indeed, most mathematically justifiable algorithms for the detection of elliptic LCSs require differentiation with respect to initial conditions \cite{Haller2015}, which is unfeasible for sparse drifter data. \\ \\
Alternatively, elliptic LCSs can be viewed as coherently evolving sets of material trajectories in space. Based on this view, a wide range of clustering methods for extracting elliptic LCSs have been proposed (see \cite{Hadjighasem2017} for a review). These objective methods differ depending on the specific clustering algorithm and the distance metric they employ to define clusters. Common algorithms include fuzzy clustering \cite{Froyland2015}, spectral graph methods \cite{Hadjighasem2016, Schlueter2017, Filippi2021} and density-based clustering methods \cite{Mowlavi2021}. Their results, however, rely on user defined input parameters such as the number of clusters, which a priori determines the possible elliptic LCS to be detected by the method in the domain. This limitation becomes even more pronounced for sparse drifters for which the number of eddies is a priori unknown. \\ \\
To extract elliptic LCS from drifter data, one is forced to rely on features of a single trajectory, such as trajectory length, velocity, acceleration and curvature. These features, however, are all inherently non-objective (frame-dependent) quantities whereas LCS, as material objects, are objective, i.e., indifferent to coordinate frame changes \cite{Haller2015}. Common single-trajectory methods, such as the absolute dispersion \cite{Provenzale1999}, trajectory length \cite{Mendoza2010, Mancho2013}, Lagrangian spin  \cite{Sawford1999}, maximal trajectory extent \cite{Mundel2014},  trajectory complexity \cite{Rypina2011} and wavelet ridge analysis \cite{Lilly2009, Lilly2010}, are limited in capturing elliptic LCSs as they are either non-objective or lack direct physical connection to material deformation in the fluid. Accordingly, while most of these methods were originally introduced to visualize LCSs from truly sparse trajectories, their application to drifter data has remained rare (see \cite{Rypina2022} for a review). Notable exceptions include the fully automated looper detection algorithms based on the Lagrangian spin \cite{Veneziani2004, Griffa2008, Dong2011, Lumpkin2016} and the wavelet ridge regression \cite{Lilly2021}. Both methods seek to extract oscillatory motions from a time-series and lead to comparable results. The methodology employed by \cite{Veneziani2004, Griffa2008, Dong2011, Lumpkin2016} is an aggregate measure of rotation within a time-series of the Lagrangian spin, whereas \cite{Lilly2021} quantifies the instantaneous oscillatory motion of the velocity using signal processing techniques. As a consequence, elliptic LCSs are visualized through looping (or oscillatory) trajectory segments. \\ \\
In order to distinguish looping from non-looping trajectory segments, a user-defined threshold is inevitably required. In a practical setting, however, differentiating between looping and non-looping trajectories proves to be challenging and hence finding an appropriate parameter is far from trivial. A common approach is to manually tune a threshold value for looping based on a subset of trajectories. Although this provides a natural and intuitive way to characterize elliptic LCSs, a considerable amount of information is lost when one discards non-looping trajectory segments. It would be more desirable to retain all trajectory information and associate to each trajectory a Lagrangian diagnostic related to material rotation in the flow. This rotational diagnostic could then be plotted over evolving trajectory positions. This approach would provide a qualitative overview of individual vortical flow structures, without relying on any chosen threshold. \\ \\
Here we propose to identify eddies from sparse drifter data using the recently introduced adiabatically quasi-objective single-trajectory diagnostics in \cite{Haller2021}. These diagnostics closely approximate appropriate objective LCS diagnostics in frames satisfying conditions that typically hold in geophysical flows. We show that the adiabatically quasi-objective trajectory rotation average ($\mathrm{\overline{TRA}}$) reveals elliptic LCSs (material eddies) at meso-and submesoscales from sparse drifter trajectory data. We additionally compare the vortical flow features extracted from the drifter-based $\mathrm{\overline{TRA}}$ computation with those obtained from other available Lagrangian single-trajectory diagnostics such as the trajectory length \cite{Mendoza2010, Mancho2013} and the Lagrangian spin \cite{Sawford1999, Lumpkin2016}. We further validate the extracted features with respect to Lagrangian averaged vorticity deviation (LAVD) \cite{Haller2016b} computed from geostrophic ocean velocity fields derived from satellite-altimetry data (AVISO).

\section{Data}

\subsection{Satellite-altimetry ocean-surface current product (AVISO)}

The two-dimensional, satellite-altimetry-derived ocean-surface current product (AVISO) has been the focus of several coherent structure studies \cite{Olascoaga2012, Olascoaga2013, Beron2015}. Using this product, the geostrophic velocity  $ \mathbf{v}_g(\mathbf{x}, t) $ of ocean currents is obtained from the remotely sensed sea-surface height $ \eta $ as
\begin{equation}
\mathbf{v}_g(\mathbf{x}, t) = \dfrac{g}{f}
\begin{pmatrix}
0 & -1 \\ 1 & 0
\end{pmatrix} \nabla \eta(\mathbf{x},t),
\end{equation}
 where $ g $ is the constant of gravity and $ f $ is the Coriolis parameter. A global, daily-gridded version of the sea-surface height profile with a spatial resolution of $ (0.25^{\circ} \times 0.25^{\circ}) $ is freely available from the Copernicus Marine Environment Monitoring Service \cite{CMEMS}.

\subsection{Drifter data sets}
While satellite-based altimetry yields mesoscale velocity fields, surface drifter observations provide direct estimates for the local surface velocity field.
In order to illustrate the range of applications of the adiabatically quasi-objective single-trajectory rotation measure introduced in \cite{Haller2021}, we will focus on two drifter data sets: the Grand Lagrangian Deployment (GLAD) and the Global Drifter Program (GDP).

\subsubsection{Grand Lagrangian Deployment (GLAD)}

We consider the Coastal Dynamics Experiment (CODE)
drifters \cite{Davis1985} released during the
Grand Lagrangian Deployment (GLAD) \cite{Olascoaga2013, Poje2014, Mariano2016}.
In order to study relative dispersion statistics, around 300 drifters were deployed on the $20^{th}-31^{th}$ of July 2012 in the northern
Gulf of Mexico, sampling various submesoscale
features over several weeks. The positions of the drifters were reported every 15 minutes from which we estimate their velocities via finite differencing. In order to highlight important circulation
features, we focus on GLAD drifters active from the $10^{th}$ to the $17^{th}$ of August, restricting the domain of interest to
\begin{equation}
\lbrace (x, y) \in [89^{\circ}\mathrm{W},86^{\circ}\mathrm{W}]\times[26^{\circ}\mathrm{N},29^{\circ}\mathrm{N}], \quad t \in [222 \ \mathrm{doy}, 229 \ \mathrm{doy}]\rbrace, 
\label{eq: GLAD-drifter}
\end{equation}
where $ \mathrm{doy} $ is in short for day of year.

\subsubsection{Global Drifter Program (GDP)}
\label{subsec: GDP}

Additionally, we consider the Global Drifter Program (GDP) data set, which contains more than 20,000 drifters released over the past 40 years. At least 1,000 of those drifters are now simultaneously active worldwide \cite{Lumpkin2007}. 
These drifters report their positions every 6 hours. We specifically focus
on a subset of drifters active from the $4^{th}$ September 2006 to the $4^{th}$ October 2006 in the Gulf Stream, i.e., in the domain
\begin{equation}
\lbrace (x, y) \in [72.5^{\circ}\mathrm{W},65^{\circ}\mathrm{W}]\times[32.5^{\circ}\mathrm{N},40^{\circ}\mathrm{N}], \quad t \in [246 \ \mathrm{doy}, 276 \ \mathrm{doy}]\rbrace.
\label{eq: GDP-drifter}
\end{equation}

\subsubsection{Drifter data preprocessing}

As higlighted by \cite{Griffa1996, Beron2016} and  \cite{Rossby2021}, inertial oscillations have little effect on the overall motion of nearby particles as they periodically return to their starting position after one inertial period. The anticyclonic looping arising from inertial oscillations, however, impacts the drifter velocity profile without affecting separation between nearby particles. In order to remove this unwanted effect from our analysis, we apply a $6^{th}$-order low-pass
Butterworth filter to the drifter trajectories with a cut-off period of $T_{cut}=1.5T_{inertial}$,
as suggested by \cite{Dong2011, Beron2016, Lumpkin2016}. Here, the inertial period is
\begin{equation}
T_{inertial} = \dfrac{2 \pi}{2 \mathrm{\omega} \sin(y)},
\label{eq: T_inertial}
\end{equation}
where $ y $ is the latitudinal position of the drifter and $ \omega = 7.27 * 10^{-5}\dfrac{\mathrm{rad}}{\mathrm{s}} $ is the rotation rate of the Earth. \newline
At mid-latitudes, inertial oscillations have time-scales of 1-2 days, whereas the dominant period of submesoscale and mesoscale eddy motion is above 5 days \cite{Mariano2016}. The characteristic time of submesoscale and mesoscale features thus greatly exceeds the inertial period. Hence, the anticyclonic looping arising from high-frequency diurnal inertial oscillations does not influence the relative dispersion of drifters in the submesoscale and mesoscale regime \cite{Beron2016}. In this paper we target coherent structures whose time-scales are far above the period of inertial oscillations.\\ \\ The time interval of definition of a finite-time, temporally aperiodic dynamical system inherently determines the range of features one can identify in that system. Submesoscale eddies have characteristic time-scales of around one week, whereas mesoscale structures  arise over months. In contrast, coherent structures tied to inertial oscillations have a temporal range of a few hours. Therefore, in the exploration of short term features, inertial oscillations should not be filtered out of the drifter trajectories.

\section{Methods}

The drifter trajectories, $\mathbf{x}(t)$, satisfy the differential
equation

\begin{equation}
\dot{\mathbf{x}}(t)=\mathbf{v}(\mathbf{x},t),\qquad \mathbf{x}(t)\in U\subset \mathbb{R}^{2},\qquad t\in[t_{0},t_{N}],\qquad \mathbf{x}(t_{0})=\mathbf{x}_{0}\label{eq: particle dynamics}.
\end{equation}
Here $\mathbf{x}$ is position, $ t $ is time and $ \mathbf{v}(\mathbf{x}(t),t) $  is the ocean-surface velocity field implied by the drifter data. While open-ocean mesoscale features are well captured by satellite-altimetry data \cite{Berta2015}, the drifter velocity $ \mathbf{v}(\mathbf{x}, t) $ in coastal regions generally differs from the geostrophic velocity component $ \mathbf{v}_g(\mathbf{x}, t) $ computed from AVISO due to coastal influences (e.g., rivers) and windage \cite{Aksamit2020}. Features derived from drifter trajectories, therefore, generally differ from those obtained from ocean geostrophic velocities. 
\subsection{Trajectory rotation average}
As already noted in the Introduction, LCSs of the velocity field are material sets and hence their existence and location are objective, i.e., indifferent to the observer. All traditionally analyzed features of trajectory data (such as length, curvature, velocity, acceleration and looping) are, in contrast, observer-dependent and hence do not allow for a self-consistent identification of LCSs. To this end, \cite{Haller2021, Haller2021Erratum} developed several quasi-objective diagnostic tools for single trajectories that do approximate objective features of trajectories in frames verifying certain conditions. Inspired by the slowly varying nature of geophysical flow data sets, we restrict our discussion to a family of frames related to each other via slowly varying (or adiabatic) Euclidean coordinate transformations
\begin{equation}
\mathbf{x} = \mathbf{Q}(t)\mathbf{y} + \mathbf{b}(t), \quad  | \mathbf{\dot{Q}} |, | \mathbf{\dot{b}} | \ll 1.
\label{eq: FrameChange}
\end{equation}
Under such slowly varying frame changes, the velocity $ \mathbf{\tilde{v}} $ transforms approximately as an objective vector, i.e,
\begin{equation}
\mathbf{\tilde{v}} = \mathbf{Q}^T(\mathbf{v}-\mathbf{\dot{Q}}^T\mathbf{y}-\mathbf{\dot{b}}) \sim \mathbf{Q}^T\mathbf{v}, 
\end{equation} where $ \mathbf{v} $ and $ \mathbf{\tilde{v}} $, respectively, denote the velocity in the original and in the slowly varying frame. We call a diagnostic adiabatically quasi-objective if in all frames related to each other via eq. (\ref{eq: FrameChange}), the diagnostic approximates the same objective quantity as long as the frames satisfy a set of additional conditions. Those conditions are specific to the quantity of interest (see, eq. (\ref{eq: A3}) and eq. (\ref{eq: Condition2}) below for examples).
\\ \\
We will apply such a single-trajectory rotation diagnostic here for the first time to actual sparse drifter data. Similarly to \cite{Haller2021}, we consider discretized drifter trajectories $\lbrace\mathbf{x}(t_{i})\rbrace_{i=0}^{N}$ satisfying eq. (\ref{eq: particle dynamics}).
As shown in \cite{Haller2021, Haller2021Erratum}, the trajectory rotation average
\begin{equation}
\mathrm{\overline{TRA}_{t_{0}}^{t_{N}}}(\textbf{x}_{0})=\dfrac{1}{t_{N}-t_{0}}\sum_{i=0}^{N-1}\cos^{-1}\dfrac{\langle\dot{\mathbf{x}}(t_{i}),\dot{\mathbf{x}}(t_{i+1})\rangle}{\vert\dot{\mathbf{x}}(t_{i})\vert\dot{\mathbf{x}}(t_{i+1})\vert}
\label{eq: TRA}
\end{equation}
approximates a time-averaged trajectory rotation relative to the overall rotation in the flow over the time interval $[t_{0},t_{N}]$, in a frame of reference satisfying
\begin{equation}
|\dfrac{\partial \mathbf{v}(\mathbf{x}(t), t)}{\partial t}| \ll |\mathbf{\ddot{x}}(t)|
\label{eq: A3}.
\end{equation}
Condition \ref{eq: A3} expresses the requirement that Lagrangian time scales dominate Eulerian time scales. This requirement has been numerically verified for the AVISO data set in \cite{Haller2021Erratum} and confirmed in several experimental studies on surface drifters \cite{Davis1985, Shepherd2000, Lumpkin2002}. As a second condition for $ \mathrm{\overline{TRA}_{t_{0}}^{t_{N}}}(\textbf{x}_{0}) $ to be an adiabatically quasi-objective scalar field (i.e. approximate an objective scalar field in qualifying frames), the Lagrangian acceleration must dominate the angular acceleration of the trajectory induced by the spatial mean vorticity:
\begin{equation}
\Bigg|\mathbf{\overline{\omega}}(t)\times \dfrac{\mathbf{\dot{x}}(t)}{|\mathbf{\dot{x}}(t)|}\Bigg| \ll \Bigg|\dfrac{d}{dt}\dfrac{\mathbf{\dot{x}}(t)}{|\mathbf{\dot{x}}(t)|}\Bigg|.
\label{eq: Condition2}
\end{equation}
This assumption has been found to hold on large enough domains in the ocean \cite{Haller2016a, Abernathey2018, Beron2019, Haller2021}. \\ \\
We associate to the trajectory $\textbf{x}(t)$ over the time interval $[t_{0},t_{N}]$ its corresponding $ \mathrm{\overline{TRA}_{t_{0}}^{t_{N}}}(\textbf{x}_{0}) $ value, where $ \mathbf{x_0}=\mathbf{x}(t_0) $. Reconstructing the $ \mathrm{\overline{TRA}_{t_{0}}^{t_{N}}}(\textbf{x}_{0}) $-field from sparse drifter trajectories via scattered interpolation allows visualization of vortical flow structures. The only parameter involved in the reconstruction of the $ \mathrm{\overline{TRA}_{t_{0}}^{t_{N}}}(\textbf{x}_{0}) $-field is the choice of the scattered interpolation method. As vortices tend to be elliptic geometric objects, we use linear radial basis function interpolation which favors such structures (see Appendix \ref{app: B} for further details). We then apply a spatial average filter of size $ (0.25^{\circ} \times 0.25^{\circ}) $ in order to suppress noise but still retain small-scale features. The size of this spatial filter coincides with the spatial resolution of the AVISO data set. Local maxima in the $ \mathrm{\overline{TRA}_{t_{0}}^{t_{N}}}(\textbf{x}_{0}) $-field mark centers of high material rotation and hence indicate underlying rotational flow features.
\subsection{Trajectory length}
\label{sec: Mfunction}

The trajectory length (also commonly referred to as the $ \mathrm{M} $-function) is a single-trajectory Lagrangian diagnostic that is applicable to sparse sets of fluid trajectories \cite{Mendoza2010, Mancho2013}. As pointed out by \cite{Ruiz2015, Ruiz2016}, this diagnostic is non-objective and has no direct relation to material stretching or rotation. 
It is nevertheless simple to compute, because the arc-length of a trajectory $ \mathbf{x}(t) $ over the time interval $ [t_0, t_N] $ starting at $ \mathbf{x}_0 $ is simply
\begin{equation}
\mathrm{M_{t_0}^{t_N}}(\mathbf{x}_0) = \sum_{i=0}^{N-1}|\mathbf{\dot{x}}(t_i)| |t_{i+1}-t_i|
\end{equation}
In an attempt to visualize elliptic LCSs from a set of sparse drifter data, we reconstruct $ \mathrm{M_{t_0}^{t_N}}(\mathbf{x}_0) $ over a broader domain using linear radial basis functions coupled with a spatial average filter of size $ (0.25^{\circ} \times 0.25^{\circ}) $, as done for the $ \mathrm{\overline{TRA}_{t_{0}}^{t_{N}}}(\textbf{x}_{0}) $ . The references \cite{Mendoza2010, Mancho2013, Mendoza2014} suggest that dynamical regions in the flow are separated by abrupt variations of $ \mathrm{M_{t_0}^{t_N}}(\mathbf{x}_0) $. Specifically, oceanic eddies are suggested to be signaled by sharp local minima of $ \mathrm{M_{t_0}^{t_N}}(\mathbf{x}_0) $.
\subsection{Lagrangian spin}
\label{sec: LagrangianSpin}

Considering looping trajectories as footprints of coherent Lagrangian eddies, \cite{Veneziani2004, Griffa2008, Dong2011, Lumpkin2016} extract looping drifter segments using the Lagrangian spin
\begin{equation}
 \mathrm{\Omega}(t_i) = \dfrac{\langle \mathbf{e}_z, \mathbf{\dot{x}}_{\mathrm{eddy}}(t_i) \times \mathbf{\ddot{x}}_{\mathrm{eddy}}(t_i)\rangle}{\dfrac{1}{2}|\mathbf{\dot{x}}_{\mathrm{eddy}}(t_i)|^2}
\label{eq: LagrangianSpin}, 
\end{equation}
first introduced in \cite{Sawford1999}. The term $ \mathbf{\dot{x}}_{\mathrm{eddy}}(t_i) $ denotes the eddy velocity and $ \mathbf{e}_z = (0,0,1)^T $. In \cite{Griffa2008}, the eddy velocity $ \mathbf{\dot{x}}_{\mathrm{eddy}}(t_i) $ was obtained by removing the background large scale flow $ \mathbf{u}_{\mathrm{avg}} $ from the velocity $ \mathbf{\dot{x}}(t_i) $ of the drifters:
\begin{equation}
\mathbf{\dot{x}}_{\mathrm{eddy}}(t_i) = \mathbf{\dot{x}}(t_i)-\mathbf{u}_{\mathrm{avg}}
\end{equation}
However, recent work points out that removing the background mean current $ \mathbf{u}_{\mathrm{avg}} $ results in misidentification of looper segments  \cite{Lumpkin2016}. Hence, we assume that the drifter velocity is dominated by the eddy fluctuating component rather than the large scale background flow:
\begin{equation}
\mathbf{\dot{x}}_{\mathrm{eddy}}(t_i) \sim \mathbf{\dot{x}}(t_i).
\end{equation}
The Lagrangian spin $ \mathrm{\Omega} $ quantifies the rotation of the velocity vector and is physically related to the vorticity. Evaluating eq. (\ref{eq: LagrangianSpin}) along a Lagrangian particle trajectory provides an overall measure of rotation within a time series over the time interval $ [t_0, t_N] $.
Looping segments of a trajectory are characterized as intervals between zero crossings of $ \Omega $. The duration of each segment of sustained positive or negative $ \mathrm{\Omega} $ is referred to as the persistence. For each segment, we define the period $ P $ as
\begin{equation}
P= \dfrac{2 \pi}{| \Omega^{*} |},
\end{equation}
where |$ \Omega^{*} $| is the absolute value of the median of $ \Omega $ over the segment. The time-series of the Lagrangian spin parameter is smoothed using a 3-day median filter in order to suppress noisy observations, as suggested by \cite{Lumpkin2016}. Looping segments are defined as trajectory intervals where the persistence exceeds the ad-hoc chosen threshold value of $ 2P $, as proposed in \cite{Lumpkin2016}. Decreasing this parameter might reveal additional looping segments, but may also lead to false positives in the identification of loopers. Hence, the identified loopers inherently depend on the choice of the minimum looping period.
\subsection{Lagrangian averaged vorticity deviation}

Similar to the $ \mathrm{\overline{TRA}} $, the Lagrangian averaged vorticity deviation (LAVD) characterizes the local material rotation in the flow relative to the overall rotation induced by the spatial mean of the vorticity \cite{Haller2016b}. Computing the LAVD value for a trajectory $ \mathbf{x}(t;\mathbf{x}_0, t_0) $ over the time interval $ [t_0,t_N] $, from the vorticity $$ \mathbf{\omega}(\mathbf{x}(t),t) = \nabla \times \mathbf{v}(\mathbf{x}(t),t) $$ gives
\begin{equation}
\mathrm{LAVD}_{t_0}^{t_N}(\mathbf{x}_0)=\dfrac{1}{t_N-t_0}\int_{t_0}^{t_N} |\mathbf{\omega}(\mathbf{x}(s),s)-\overline{\mathbf{\omega}}(s)|ds,
\end{equation}
where $ \overline{\mathbf{\omega}}(t) $ denotes the spatially averaged vorticity at time $ t $.
Therefore, even though the LAVD is a quantity associated with a single trajectory, its computation requires the knowledge of the velocity field over a large enough domain. This renders the LAVD diagnostics inapplicable to sparse drifter data, even though it has been used to visualize and extract vortices from gridded ocean velocity data sets \cite{Haller2016b, Abernathey2018, Beron2020}. Local maxima in the $ \mathrm{LAVD} $-field surrounded by nested, closed and nearly convex level curves highlight elliptic LCSs.
In this work, we apply the LAVD diagnostic to the AVISO data and compare the eddies obtained in this way with those revealed by the $ \mathrm{\overline{TRA}} $ from drifter data.

\subsection{Validation of eddy boundary extraction algorithm on numerical ocean model}
\begin{figure}[http]
\centering \includegraphics[scale=0.7]{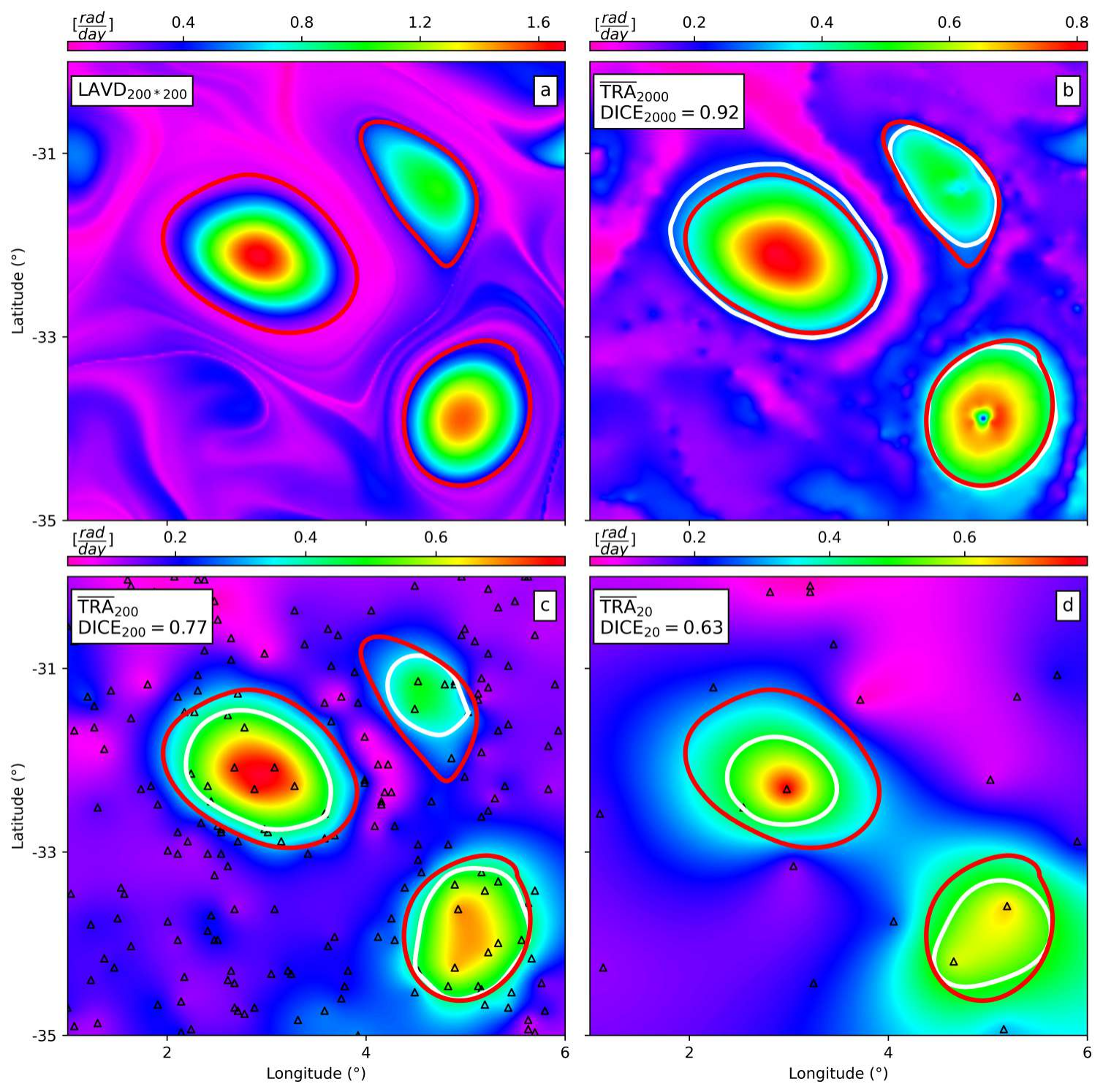} \caption{ \label{fig:LAVD_TRA} $\mathrm{LAVD} $ and $ \mathrm{\overline{TRA}}$ evaluated on the full and progressively subsampled AVISO data sets in the Agulhas region. (a) $ \mathrm{LAVD} $-field with full resolution ($ 200 \times 200$). The red curves denote the boundaries of the three $ \mathrm{LAVD} $-eddies. (b-d) Reconstructed $\mathrm{\overline{TRA}}$-field from $ 2000,200 $ and $ 20 $ randomly selected trajectories. The white curves indicate boundaries of the $ \mathrm{\overline{TRA}} $-eddies. The black triangles indicate final trajectory positions.}
\end{figure}

We first validate the eddy boundary extraction algorithm proposed in Appendix \ref{app: A} on a numerical ocean model obtained from the AVISO surface velocity field \cite{CMEMS}. In our text, we focus on the Agulhas region, which contains a variety of mesoscale eddies analyzed previously by several coherent structure studies \cite{Haller2016a, Haller2021}. We specifically consider the area of the Agulhas leakage,
\begin{equation}
U = \lbrace (x,y) \in [1,6] \times [-35,-30] \rbrace,
\end{equation}
over a period of 25 days. We first compute the $ \mathrm{LAVD} $ from $ 200 \times 200$ densely gridded trajectories and then successively and randomly subsample this trajectory set for a comparative calculation of the $ \mathrm{\overline{TRA}} $. The $ \mathrm{LAVD} $ highlights three mesoscale eddies (see Fig. (\ref{fig:LAVD_TRA}a)), whose boundaries are obtained using the algorithm of \cite{Haller2016a} with the convexity deficiency set to $ 10^{-3} $.
 These three eddies are clearly visible in the $  \mathrm{\overline{TRA}} $-field (Fig. (\ref{fig:LAVD_TRA}b-c)), even after a progressive subsampling of the trajectory density to $ \dfrac{200}{5^{\circ} \times 5^{\circ}} $. Upon a further decrease of the trajectory density to $ \dfrac{20}{5^{\circ} \times 5^{\circ}} $, two out of the three mesoscale features still persist in the $ \mathrm{\overline{TRA}} $-field (see Fig. (\ref{fig:LAVD_TRA}d)). The third mesoscale feature is not captured by the $ \mathrm{\overline{TRA}} $-field only because no remaining trajectory samples that region after the random sparsification we applied to the original trajectory data set. \\ \\
We now quantitatively compare the $ \mathrm{\overline{TRA}} $-eddies with the $ \mathrm{LAVD} $-eddies using the DICE coefficient that measures the area overlap between the corresponding eddies \cite{Lguensat2018}. We denote the domain covered by the $ \mathrm{\overline{TRA}} $-eddies by $ \mathrm{Area(\overline{TRA}_{N, eddy})} $, where $ N $ is the number of trajectories used to reconstruct the $ \mathrm{\overline{TRA}} $-field. $ \mathrm{LAVD} $-eddies that contain at least one trajectory are taken as groundtruth for eddy detection with their domain denoted by $ \mathrm{Area(LAVD_{eddy})} $. The DICE coefficient for eddy detection comparison is then defined as
\begin{equation}
\mathrm{DICE_N} = \dfrac{2 (\mathrm{Area(\overline{TRA}_{N, eddy})} \cap \mathrm{Area(LAVD_{eddy})})}{\mathrm{Area(\overline{TRA}_{N, eddy})}+\mathrm{Area(LAVD_{eddy})}}.
\label{eq: DICE}
\end{equation} A DICE coefficient of 1 would indicate perfect overlap whereas 0 would indicate no overlap between $ \mathrm{LAVD} $-eddies and $ \mathrm{\overline{TRA}} $-eddies.
\begin{figure}[http]
\centering \includegraphics[scale=0.6]{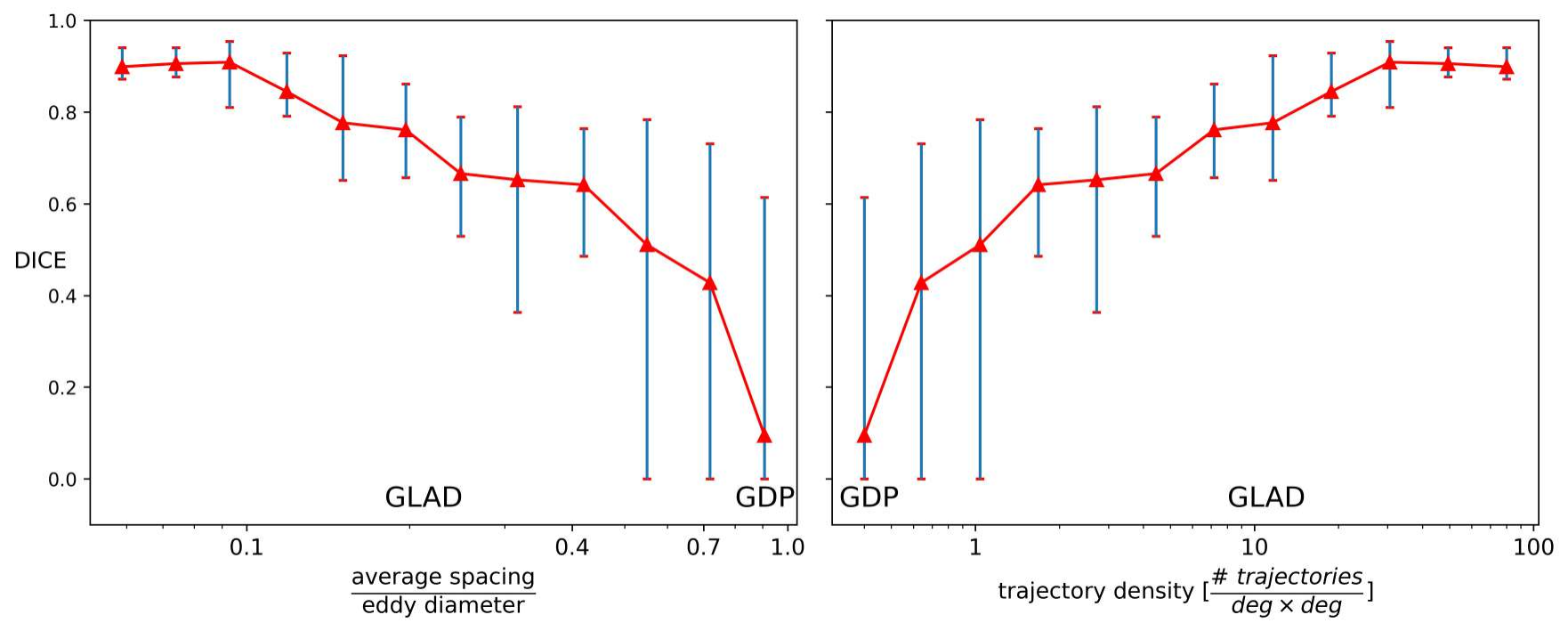} \caption{ \label{fig:DICE} Averaged DICE coefficient plotted as a function of the averaged trajectory spacing per eddy diameter. (a) and trajectory density (b). The error bars indicate the minimum and maximum DICE coefficient attained within the corresponding set of trajectories.}
\end{figure} \\ \\
We vary the number of trajectories from $ 10 $ to $ 2000 $ and randomly subsample multiple times in order to obtain a statistical estimate of the DICE coefficient for each trajectory density. In Fig. (\ref{fig:DICE}), we plot the averaged DICE coefficient and its error bar as a function of two quantities: the averaged trajectory spacing normalized by the eddy diameter (Fig. (\ref{fig:DICE}a)) and the trajectory density (Fig. (\ref{fig:DICE}b)). The eddy diameter of the three $ \mathrm{LAVD}$-eddies in the Agulhas region is around $ 100 \mathrm{km} \ (\sim 1^{\circ}) $. The averaged trajectory spacing per eddy diameter relates the averaged drifter spacing to the eddy length scales we seek to resolve. Decreasing the drifter density (or, equivalently, increasing the averaged spacing) leads to a drop in the DICE coefficient. Even for very sparse data sets (e.g. $ \dfrac{\mathrm{average \ spacing}}{\mathrm{eddy \ diameter}} \sim 1 $), it is possible to approximately identify coherent structures with maximum DICE coefficients of around $ 0.6 $. Note, however, that the results inevitably depend on the drifter distribution. If no drifters are inside an eddy, then we do not expect the eddy to be visible in the $ \mathrm{\overline{TRA}} $-field.  \\ \\
GLAD drifters (to be analyzed in section \ref{subsec: GLAD}) have an average spacing of $ 5 \mathrm{km} $ and a trajectory density of around $ \dfrac{10}{\mathrm{deg} \times \mathrm{deg}} $. The diameter of the smallest submesoscale eddies observed during the GLAD experiment is around $ 25 \ \mathrm{km} $ and the averaged drifter spacing per eddy diameter is $ 0.2 $. GDP drifters (to be analyzed in section \ref{subsec: GDP_example}) are on average separated by $ 150 \ \mathrm{km}$ with an estimated trajectory density of $ \dfrac{0.2}{\mathrm{deg} \times \mathrm{deg}} $. Mesoscale eddies have characteristic widths ranging from $ 100 $ to $ 300 \ \mathrm{km} $ \cite{Lumpkin2016} and hence the averaged spacing per eddy diameter ranges from $ 0.5 $ to $  1.5 $. Based on the statistical estimates of the DICE coefficient in Fig. (\ref{fig:DICE}), we expect to detect submesoscale and mesoscale eddies in the GLAD and GDP data sets with reasonable confidence. 
\section{Results}
\subsection{Grand Lagrangian Deployment (GLAD)}
\label{subsec: GLAD}
\begin{figure}[http]
\centering \includegraphics[scale=.73]{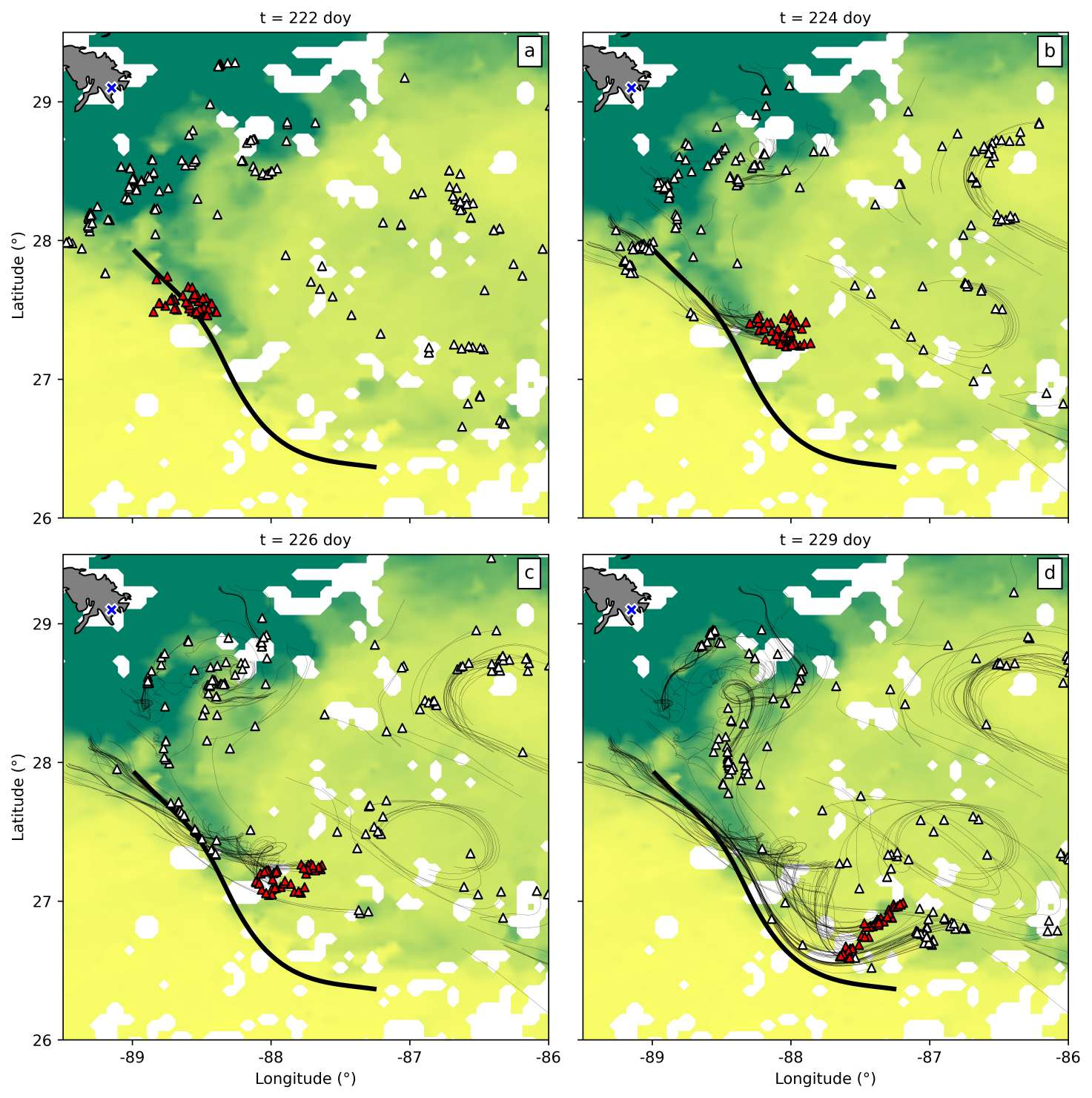} \caption{ \label{fig:GLAD_Overview} Evolution of GLAD-drifters overlayed on the 8-day composite chlorophyll concentration (12 August, 2012). Most drifters (white triangles) formed a long filament along the chlorophyll plume (attracting LCS) extending deep into the Gulf of Mexico. Drifters additionally forming a visible tight cluster are colored red. The blue cross denotes the outlet of the Mississippi river. The black line corresponds to the attracting LCS. (a) $t=222 \ \mathrm{doy}$. (b) $t=224 \ \mathrm{doy}$. (c) $t=226 \ \mathrm{doy}$. (d) $t=229 \ \mathrm{doy}$.}
\end{figure}

Drifters released in the GLAD experiment have mostly been used to study dispersion in the ocean over a range of scales \cite{Poje2014, Beron2016, Mariano2016}. Furthermore, refs. \cite{Olascoaga2012} and \cite{Olascoaga2013} found correlation between the evolution of the drifters and nearby attracting LCSs extracted from the geostrophic velocity field. Specifically, drifter behavior agreed with the \textit{tiger tail} pattern inferred from the chlorophyll distribution shown in Fig. (\ref{fig:GLAD_Overview}). A chlorophyll plume extended over more than 100 kilometres from the outlet of the Mississippi river into the open sea and coincided with an attracting LCS \cite{Olascoaga2013}. The attracting LCS (continuous black line in Fig. (\ref{fig:GLAD_Overview})) is computed from the geostrophic velocity field $ \mathbf{v_{\mathrm{g}}} $ using backward trajectories over the time interval $ [222 \ \mathrm{doy}, 229 \ \mathrm{doy}] $ according to \cite{Farazmand2012}.
 Drifters in the proximity of the tip of the tiger tail organized themselves into long filaments along the attracting LCS. Some of the drifters (in red) additionally exhibited some clustering, suggesting the presence of an elliptic LCS close to the chlorophyll front.\\ \\
\begin{figure}[http]
\centering \includegraphics[scale=.71]{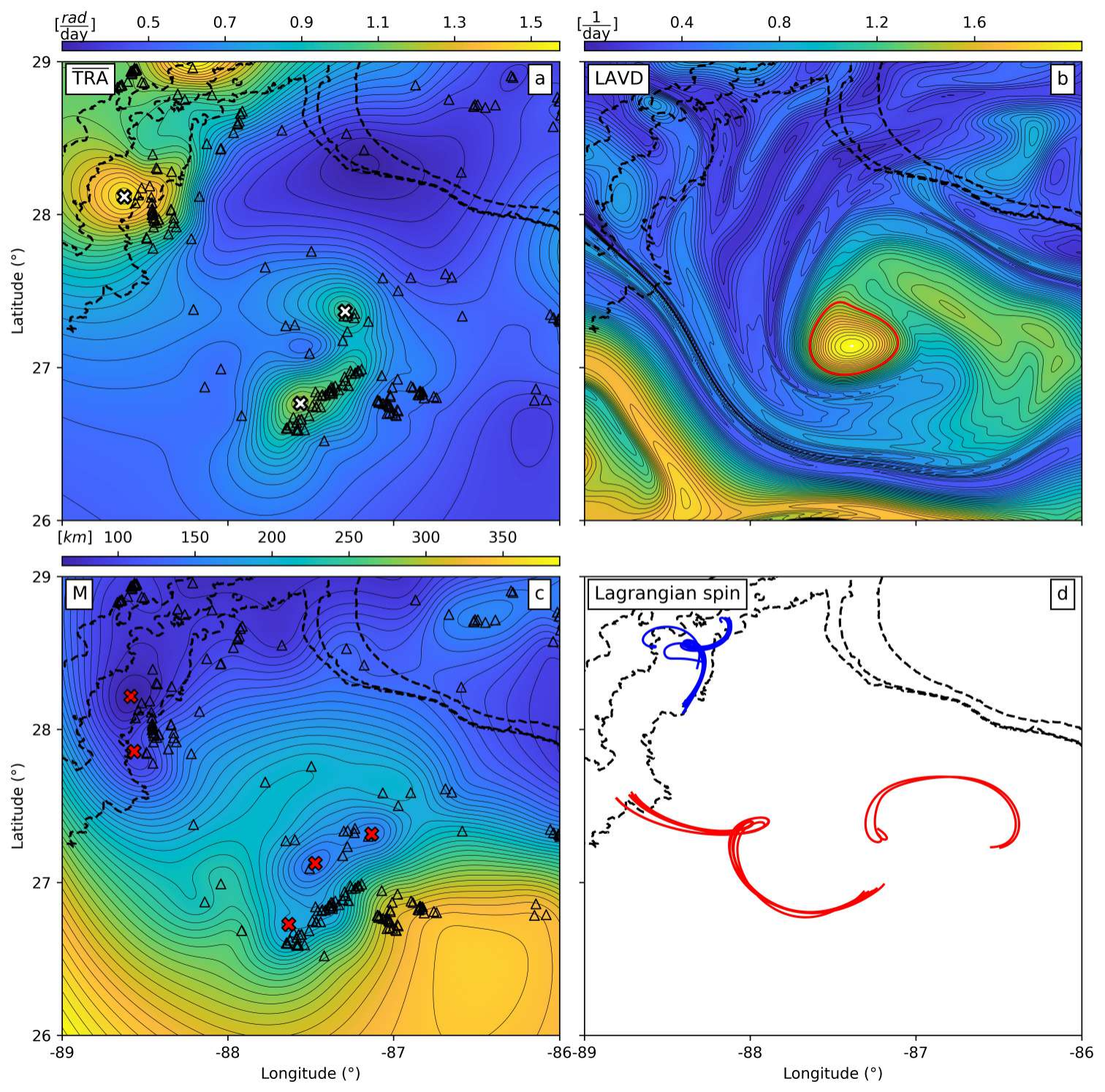} \caption{ \label{fig:TRA_GLAD_end} Different material eddy diagnostics computed from the GLAD data set or the AVISO data set. The dashed black lines correspond to the isobaths $ (-1000 \ \mathrm{m}, -1500 \ \mathrm{m}, -2000 \ \mathrm{m}) $ and highlight the coastal area. (a) Reconstructed $ \mathrm{\overline{TRA}_{222}^{229}} $-field plotted with respect to the position of the drifters at time $ t = 229 \ \mathrm{doy}$. White crosses indicate local maxima. (b) $ \mathrm{LAVD_{222}^{229}} $-field computed from AVISO plotted at time $ t = 229 \ \mathrm{doy}$. The closed red curve denotes the eddy boundary extracted from the $ \mathrm{LAVD}_{222}^{229} $-field.  (c) Reconstructed $ \mathrm{M_{222}^{229}} $-field plotted with respect to the position of the drifters at time $ t = 229 \ \mathrm{doy}$.  Red crosses indicate local minima. (d) Cyclonic (red) and anticyclonic (blue) loopers.}
\end{figure}Here we seek to visualize vortices by reconstructing the $ \mathrm{\overline{TRA}} $-field from drifter data over the interval $ [222 \  \mathrm{doy}, 229 \ \mathrm{doy}] $ using linear radial basis function interpolation. The $ \mathrm{\overline{TRA}} $, shown in Fig. (\ref{fig:TRA_GLAD_end}), is plotted with respect to the final drifter positions at time $ t = 229 \ \mathrm{doy} $. This plot reveals multiple rotational features marked by local maxima of the $ \mathrm{\overline{TRA}} $ surrounded by a dense set of closed and almost convex contours. For comparison, we include three further Lagrangian eddy diagnostics: the LAVD computed from geostrophic velocity currents, the trajectory length (also called $ \mathrm{M} $-function) and the Lagrangian spin computed from drifter data.

\subsubsection{Coastal area}
We start by focusing on the area at the outlet of the Mississippi river on the continental shelf. Coastal flows are regions of intense mixing \cite{Aksamit2020}, often characterized by high potential vorticity and strong horizontal shear, which are responsible for the formation of small-scale eddies \cite{Southwick2016, Southwick2017}. Anticyclonic looping segments extracted from the Lagrangian spin parameter, confirm the presence of small-scale vortices at the outlet of the Mississippi river (Fig. (\ref{fig:TRA_GLAD_end}d)). Evidence for the existence of submesoscale elliptic features and strong mixing areas on the continental shelf is also found in the $ \mathrm{\overline{TRA}} $-field (Fig. (\ref{fig:TRA_GLAD_end}a)). Vortices are revealed in the $ \mathrm{\overline{TRA}} $ as local maxima surrounded by a dense set of closed curves indicating abrupt spatial variations. The local maximum of the $ \mathrm{\overline{TRA}} $ at $ (88.6^{\circ}W, 28.1^{\circ}N) $ is surrounded by a dense set of closed contours, thereby indicating high spatial gradients in the $ \mathrm{\overline{TRA}} $-field.
Close to the vortical flow features identified by the $ \mathrm{\overline{TRA}} $ in the coastal area, the $ \mathrm{M} $-function displays two local minima, respectively at $ (88.6^{\circ}W, 28.2^{\circ}N) $ and $ (88.5^{\circ}W,27.8^{\circ}N) $ (Fig. (\ref{fig:TRA_GLAD_end}c)). Similarly to the $ \mathrm{\overline{TRA}} $ and the Lagrangian spin parameter, the $ \mathrm{M} $-function also indicates the existence of coastal eddies. However, compared to the $ \mathrm{\overline{TRA}} $, which displays a unique local maximum surrounded by a dense set of closed and convex curves, the $ \mathrm{M} $-function displays multiple local minima surrounded by sparse level sets, thereby indicating a region without sharp minima.
\\ \\
The $ \mathrm{LAVD} $-field displays no distinguished features close to the outlet of the Mississippi river (Fig. (\ref{fig:TRA_GLAD_end}b)). Eddy boundaries can be selected as outermost (almost) convex contours around a local maximum in the $ \mathrm{LAVD} $-field \cite{Haller2016b}. Passing from the $ \mathrm{LAVD} $-field to a discrete set of closed curves requires specifying two parameters: the minimum length $ \mathrm{l_{min}} $ of the perimeter of the eddy and the convexity deficiency $ \mathrm{c_d} $. The convexity deficiency is
\begin{equation}
\mathrm{c_d} = \dfrac{|\mathrm{A_{contour}-A_{contour, convex}}|}{\mathrm{A_{contour}}},
\end{equation}
where $ \mathrm{A_{contour}} $ is the area enclosed by the closed contour and $ \mathrm{A_{contour,convex}} $ is the area enclosed by the convex hull of the closed contour. Similarly to \cite{Haller2016b}, we set $ \mathrm{c_d} = 10^{-3} $. As a threshold for the perimeter of the eddy $ \mathrm{l_{min}} $, we select $ \mathrm{l_{min}} = 2\pi \mathrm{r}$, with $ \mathrm{r} = 25\mathrm{km} \ (\sim 0.25^{\circ}) $. This corresponds to the minimal length scales that can reliably be resolved by the AVISO data set with a spatial resolution of $ (0.25^{\circ}\times 0.25^{\circ})  $. The $ \mathrm{LAVD} $ does not reveal the presence of coastal eddies (Fig. (\ref{fig:TRA_GLAD_end}b)). We attribute the mismatch between the $ \mathrm{LAVD} $ and the above discussed drifter-based diagnostics to the insufficient resolution of the underlying ocean surface velocity close to the outlet of the Mississippi river by the AVISO data set. Hence, in coastal areas, the drifter-based $ \mathrm{\overline{TRA}} $ diagnostic captures the flow dynamics with higher detail than the AVISO-based $ \mathrm{LAVD} $.
\subsubsection{Open sea}
The $ \mathrm{\overline{TRA}} $ highlights another family of elliptic LCSs  close to the chlorophyll plume extending from the outlet of the Mississippi river to the open ocean (Fig. (\ref{fig:TRA_GLAD_end}a)). The two local maxima marking elliptic LCSs are located at $  (87.5^{\circ}W, 26.8^{\circ}N) $ and $ (87.3^{\circ}W, 27.4^{\circ}N) $. Due to the close proximity of these extrema, it is, however, a priori unclear whether they highlight two separate eddies or whether they are part of the same mesoscale eddy. The cyclonic looping segments confirm the presence of vortical flow features in this area (red trajectories in Fig. (\ref{fig:TRA_GLAD_end}d)), but do not specifically highlight eddy boundaries. Inspection of the cyclonic trajectories suggests that they are originally part of two distinct eddies as they come from two separate regions. Extracting looping trajectory segments from the time-series of the Lagrangian spin is a powerful methodology. However, discriminating between loopers and non-loopers requires an ad-hoc choice of the minimum sustained looping period of a trajectory. In contrast, the $ \mathrm{\overline{TRA}} $ retains all trajectory information and allows visualization of vortical flow structures from sparse drifter data using a scalar diagnostic.
\\ \\
The investigated elliptic LCSs highlighted by the $ \mathrm{\overline{TRA}} $ and the Lagrangian spin are also visible in the $ \mathrm{LAVD} $, which displays a nearly convex vortical flow feature (closed red curve in Fig. (\ref{fig:TRA_GLAD_end}b)). Compared to the $ \mathrm{\overline{TRA}} $, however, which indicates two separate albeit closely located local maxima, the $ \mathrm{LAVD} $ computation clearly reveals a single mesoscale eddy. \\ \\
In contrast to the aforementioned methods, the features resulting from the trajectory length diagnostic are far less pronounced (Fig. (\ref{fig:TRA_GLAD_end}c)). Indeed, the $ \mathrm{M}$-function displays multiple local minima, which are only partially correlated with the eddies suggested by the other methods. Furthermore, local minima in the $ \mathrm{M} $-function are not sharp, as indicated by the sparsity of the surrounding level curves. Hence, extracting eddy-related features from the $ \mathrm{M} $-function is challenging as there exist no distinguishable sharp and closed boundaries surrounding the local minima.
\subsubsection{Eddy dynamics}
\label{sec: QMA}
\begin{figure}[http]
\centering \includegraphics[scale=0.72]{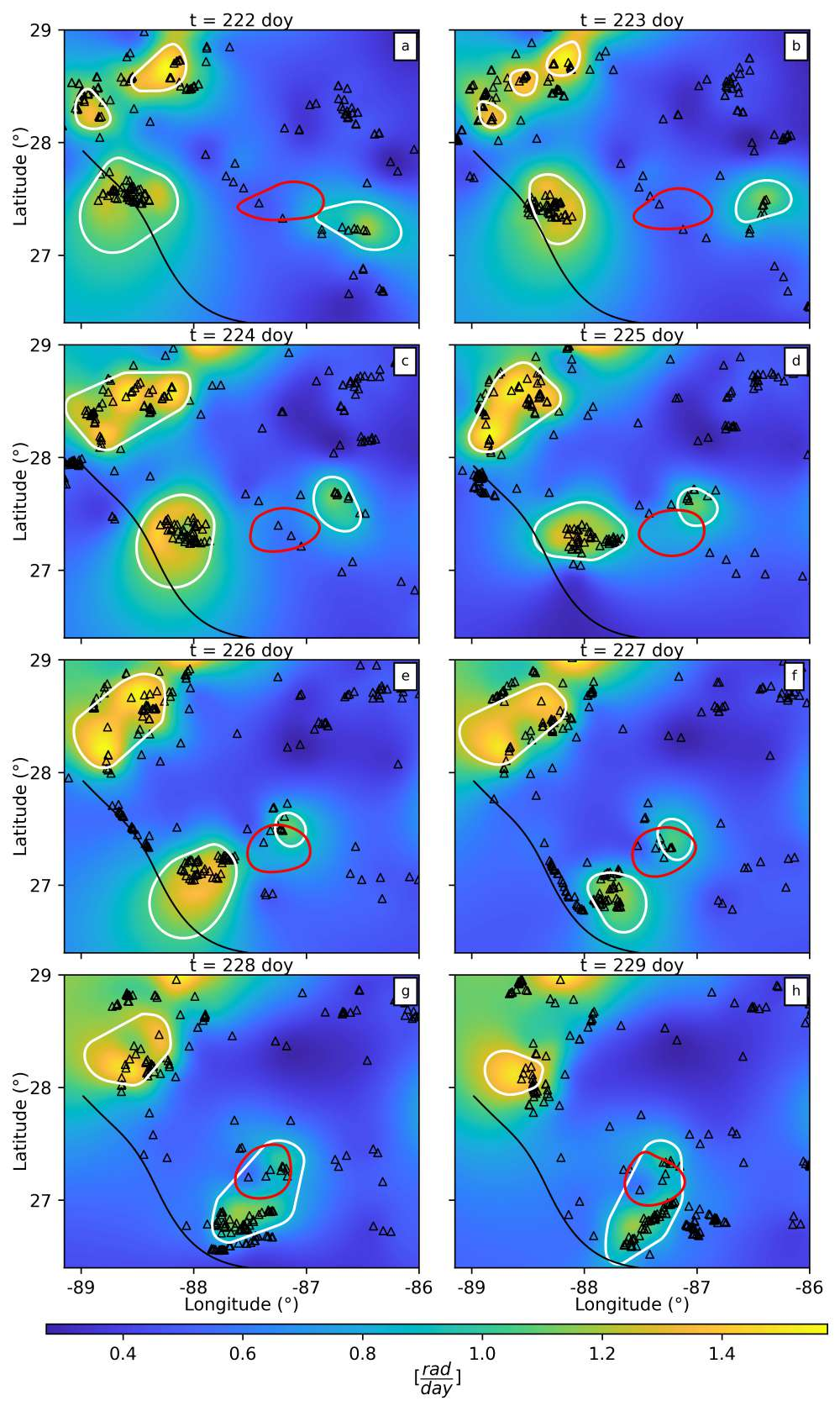} \caption{ \label{fig:TRA_drifter_evolution_GLAD} Quasi-materially advected $ \mathrm{\overline{TRA}_{222}^{229}}$ for the GLAD data set over t $ \in $ [222 doy, 229 doy] (a-h). The black line denotes the AVISO-based attracting LCS. Closed white curves mark eddy boundaries extracted from the drifter-based $ \mathrm{\overline{TRA}} $ using the algorithm described in Appendix \ref{app: A}. Closed red curves indicate eddy boundaries obtained from the AVISO-based $ \mathrm{LAVD} $ computation. Triangles indicate the positions of the drifters.}
\end{figure}

\begin{figure}[http]
\centering \includegraphics[scale=0.7]{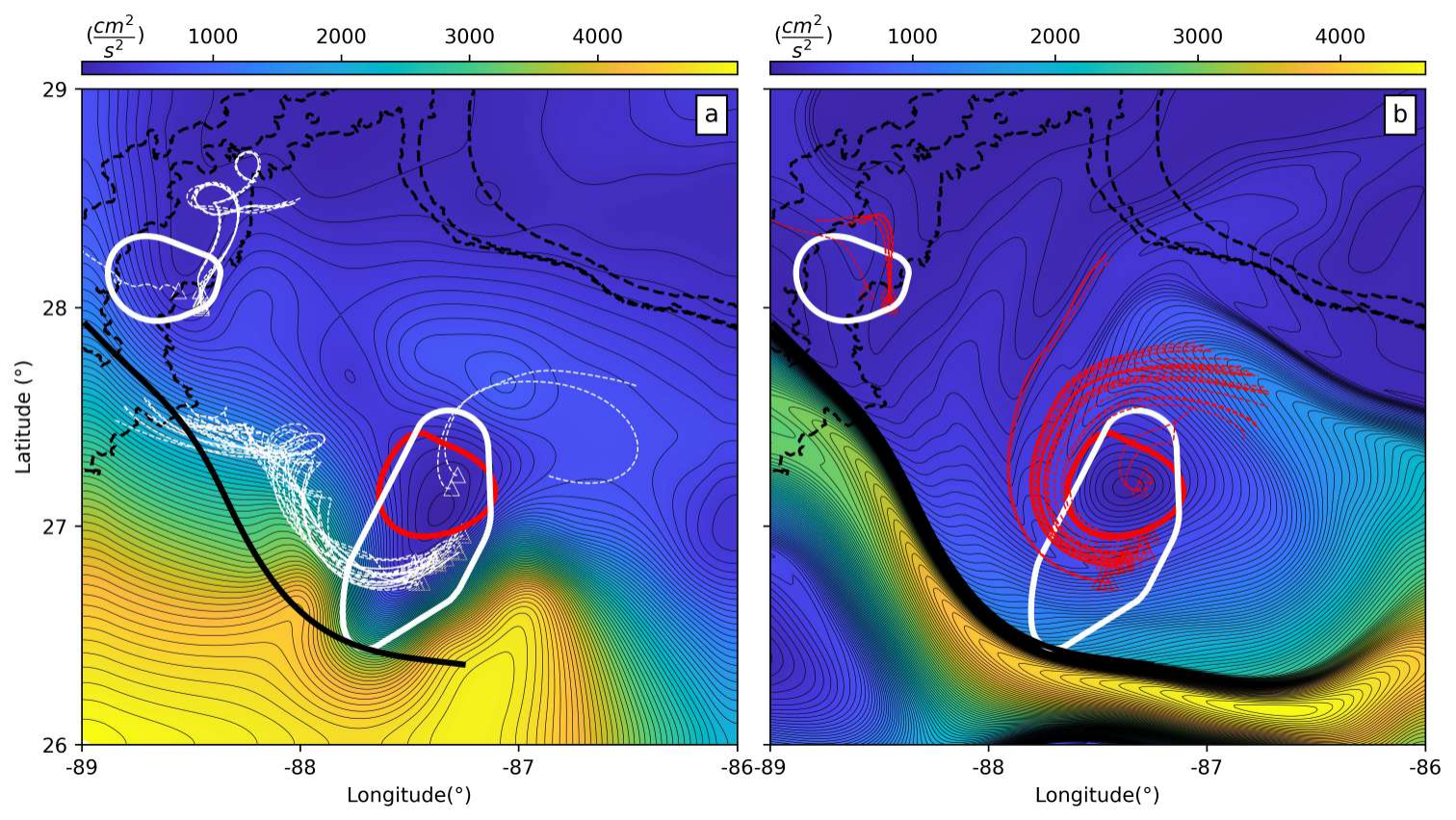} \caption{ \label{fig:KE} $ \mathrm{\overline{KE}} $ evaluated both on GLAD-drifters and  on AVISO. The white and red eddies respectively indicate the eddies extracted from the drifter-based $ \mathrm{\overline{TRA}} $-field and the AVISO-based $ \mathrm{LAVD} $-field. The dashed black lines correspond to the isobaths $ (-1000 \ \mathrm{m}, -1500 \ \mathrm{m}, -2000 \ \mathrm{m}) $. The black line denotes the AVISO-based attracting LCS. (a) Drifter-based $ \mathrm{\overline{KE}_{222}^{229}} $-field at time $ t = 229 \ \mathrm{doy} $. White, dashed curves denote a subset of GLAD drifter trajectories contained currently in the same $ \mathrm{\overline{TRA}} $-eddy. (b) AVISO-based $ \mathrm{\overline{KE}_{222}^{229}} $-field at time $ t = 229 \ \mathrm{doy} $. Red, dashed curves denote a subset of AVISO trajectories contained currently in the same $ \mathrm{\overline{TRA}} $-eddy.}
\end{figure}

In order to investigate the formation and evolution of the elliptic LCSs inferred from the drifter-based $ \mathrm{\overline{TRA}} $, we proceed by quasi-materially advecting the $ \mathrm{\overline{TRA}} $ distribution over the time interval $ [222 \  \mathrm{doy}, 229 \ \mathrm{doy}] $ (Fig. (\ref{fig:TRA_drifter_evolution_GLAD})). True material advection would require a spatiotemporally well-resolved velocity field. In our setting, however, the velocity field is only sparsely known and hence the advected structures are inherently non-material: At every time step, the $ \mathrm{\overline{TRA}} $-field must be approximated from the current drifter distribution. As a consequence, the evolution of the extracted eddy boundaries is not exactly Lagrangian. This implies that a set of drifters may not stay inside an eddy over the full time interval. For the same reason, the extracted eddies can potentially merge. \\ \\
The closed white curves in Fig. (\ref{fig:TRA_drifter_evolution_GLAD}) indicate eddy boundaries extracted from the $ \mathrm{\overline{TRA}} $ using the algorithm described in Appendix \ref{app: A}. The red closed curve denotes the truly materially advected vortex boundary extracted from the $ \mathrm{LAVD} $ at time $ \mathrm{t}=229 \ \mathrm{doy} $. The $ \mathrm{LAVD} $-based eddy at $ t = 229 \ \mathrm{doy} $ is materially advected using the geostrophic velocity $\mathbf{v}_g(\mathbf{x},t) $, whereas the eddy boundaries inferred from the drifter-based $ \mathrm{\overline{TRA}} $ are quasi-materially advected along drifter trajectories. \\ \\ 
At $ t = 222 \ \mathrm{doy} $, the $ \mathrm{\overline{TRA}} $ suggests the presence of several small-scale vortices at the outlet of the Mississippi river and at open sea (Fig. (\ref{fig:TRA_drifter_evolution_GLAD}a)). The submesoscale eddies close to the outlet of the Mississippi river remain trapped in coastal areas and eventually merge into a larger vortical flow feature. Over the time interval $ [222 \ \mathrm{doy}, 226 \ \mathrm{doy}] $, the $ \mathrm{LAVD} $-based eddy does not coincide with any of the eddies inferred from the $ \mathrm{\overline{TRA}}$ (Fig. (\ref{fig:TRA_drifter_evolution_GLAD}a-d)). The white eddy initially located at approximately $ (88.5^{\circ}W, 27.5^{\circ}N) $ is associated with the clustered red drifters identified in Fig. (\ref{fig:GLAD_Overview}), thereby confirming the existence of the submesoscale eddy along the chlorophyll front, which agrees with the observation put forward in \cite{Olascoaga2012, Olascoaga2013}. The eddy develops along the attracting LCS and then slowly detaches away from the oceanic front (Fig. (\ref{fig:TRA_drifter_evolution_GLAD}a-f)). Drifters are thereby coherently transported from the coastline into the open sea. The drifter-based eddy evolving along the AVISO-based attracting LCS eventually merges with the submesoscale eddy originally located at $ (86.5^{\circ}W, 27.25^{\circ}N) $ to form a larger mesoscale eddy at $ \mathrm{t} = 228 \ \mathrm{doy} $ (Fig. (\ref{fig:TRA_drifter_evolution_GLAD}g)). Hence, elliptic LCSs generated along oceanic fronts offer a transport mechanism for particles, carrying material over long distances away from the coastline into the open sea. We point out that intersections between attracting and elliptic LCS (such as at $ t = 222 \mathrm{doy} $) are physically inconsistent as they imply contradicting material response. We attribute this inconsistency to the fact that the attracting LCS is computed from $ \mathbf{v}_g(\mathbf{x},t) $ whereas the white elliptic LCSs are computed from drifter velocities. \\ \\
Towards the end of the advection process, the $ \mathrm{LAVD} $-based elliptic LCS (red curve in Fig. (\ref{fig:TRA_drifter_evolution_GLAD}h)) approximates the mesoscale eddy inferred from the $ \mathrm{\overline{TRA}} $ (white curve centered at approximately $ (87.5^{\circ}W, 27^{\circ}N) $ in Fig. (\ref{fig:TRA_drifter_evolution_GLAD}h)). The red eddy shows no degree of filamentation and remains coherent over the full time interval. The white eddy results from the vortex merger between two smaller eddies and is significantly larger than the red eddy. At $ \mathrm{t}=229 \ \mathrm{doy} $, the circumference of the $ \mathrm{\overline{TRA}} $-based white eddy and the $ \mathrm{\overline{LAVD}} $-based red eddy are $ 298.5 \mathrm{km} $ and $ 152 \mathrm{km} $. As the $ \mathrm{LAVD} $-based eddy is a materially advected closed curve, it can neither split nor merge with any other materially advected curve. Hence, by construction, the $ \mathrm{LAVD} $ is unable to capture vortex mergers. In contrast, the quasi-materially advected, $ \mathrm{\overline{TRA}} $-based eddies can merge into larger eddies. This follows because the evolving eddies are not purely Lagrangian, given that the computation of the eddy boundary from the $ \mathrm{\overline{TRA}} $-field is independently carried out at each time step.
 \\ \\
The kinetic energy (KE) is frequently seen as an indicator of large-scale coherent motion in the ocean \cite{Richardson1983, Lumpkin2013, Lumpkin2016}.
The averaged KE associated to a trajectory $ \mathbf{x}(t) $ over the time interval $ [t_0, t_N] $ starting at $ \mathbf{x}_0 $ is
\begin{equation}
\mathrm{\overline{KE}_{t_0}^{t_N}}(\mathbf{x}_0) = \dfrac{1}{2|t_N-t_0|}\sum_{i=0}^{N-1}|\mathbf{\dot{x}}(t_i)|^2 |t_{i+1}-t_i|.
\end{equation}
$ \mathrm{\overline{KE}} $ is a non-objective, Lagrangian single-trajectory diagnostics that bears similarities to the trajectory length diagnostic introduced in section (\ref{sec: Mfunction}). It is often used to visualize energetically dominant flow structures both from AVISO \cite{Martinez2019} and drifter data \cite{Lumpkin2016}. Here we compute the AVISO-based $ \mathrm{\overline{KE}} $-field from synthetic Lagrangian particle trajectories generated by the geostrophic velocity field $ \mathbf{v}_g(\mathbf{x}, t) $ in the region of interest (Fig. (\ref{fig:KE}b)). The computed $ \mathrm{\overline{KE}} $-field displays a front-like feature coinciding with the attracting LCS responsible for the transport drifters from the outlet of the Mississippi river into the open sea. The drifter-based $ \mathrm{\overline{KE}} $-field, reconstructed from drifter data shows a front-like feature reminiscent of the front visible in the AVISO-based $ \mathrm{\overline{KE}} $ (Fig. (\ref{fig:KE}a)). The weakly spiraling feature in the AVISO-based $ \mathrm{\overline{KE}} $-field corresponds to the mesoscale elliptic LCSs. Overall, however, both methodologies fail to clearly highlight the oceanic eddies detected from $ \mathrm{\overline{TRA}} $ and $ \mathrm{LAVD} $. Hence, although frequently related to coherent eddy motions, we conclude that the kinetic energy is not conclusively linked to oceanic eddies in the region of interest.
 Figure (\ref{fig:KE}) also shows drifters (white) and AVISO trajectories (red) released within $ \mathrm{\overline{TRA}} $-based eddies.  Overall, the real and synthetic drifter trajectories divert from each other over time, as expected. The difference is more pronounced in coastal areas due to the influence of external factors such as the outflow of the Mississippi-river.
\\ \\
The quasi-material advection of the $ \mathrm{\overline{TRA}} $-based eddies sheds new light on the origin and the formation of the mesoscale eddy detected from the AVISO-based $ \mathrm{LAVD} $-field. Despite the minimal amount of data, the $ \mathrm{\overline{TRA}} $ reveals eddy dynamics hidden in the $ \mathrm{LAVD} $ and $ \mathrm{\overline{KE}} $.

\subsection{Global Drifter Program (GDP)}
\label{subsec: GDP_example}
\begin{figure}[http]
\centering \includegraphics[scale=0.73]{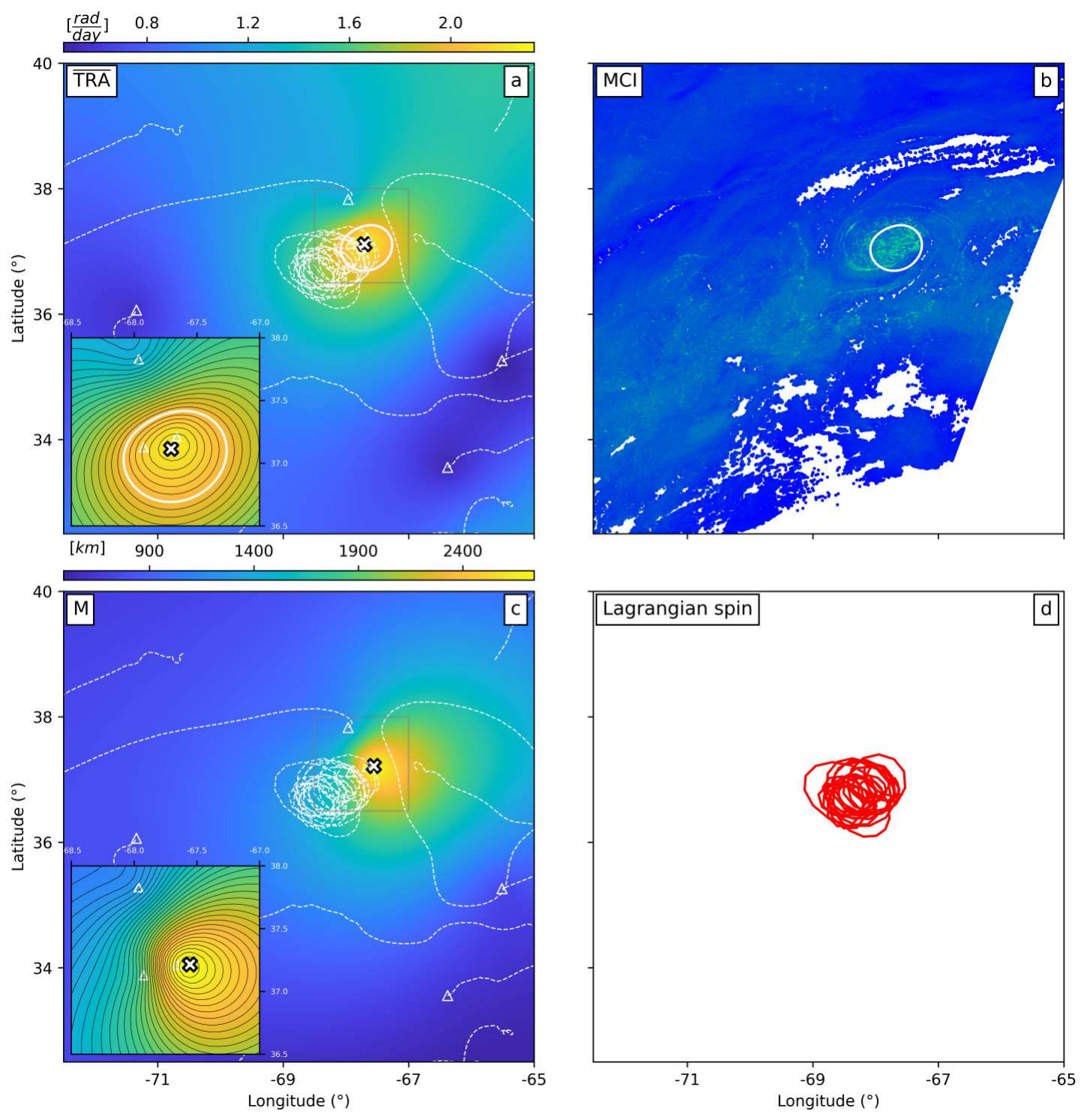} \caption{ \label{fig:Sargassum} Lagrangian diagnostics evaluated on a subset of the GDP data set. (a) Reconstructed $ \mathrm{\overline{TRA}_{246}^{276}} $-field plotted with respect to the position of the drifters (triangles) at time $ t = 276 \ \mathrm{doy} $. White cross indicates unique local maximum in the $ \mathrm{\overline{TRA}_{246}^{276}} $-field. (b) Eddy boundary superimposed on the floating sargassum concentration inferred from the Maximum Chlorophyl Index (MCI) at $ t = 276 \ \mathrm{doy} $. (c) Reconstructed $ \mathrm{M_{246}^{276}} $-field plotted with respect to the position of the drifters (triangles) at time $ t = 276 \ \mathrm{doy} $. White cross indicates unique local maximum in the $ \mathrm{M_{246}^{276}} $-field. (d) Cyclonic (red) loopers.}
\end{figure}

In our second example, we focus on a set of drifters in the western North-Atlantic. This oceanic region is characterized by a strong and persistent formation of eddies arising from the
meanders of the Gulf Stream \cite{Kang2013, Richardson1973}.
On the $ 4^{th} $ of October, 2006, a floating sargassum patch was detected by the Medium Resolution Imaging Spectrometer (MERIS) on Envisat. This feature has a spiralling shape that is also visible from satellite-altimetry data \cite{Beron2015}. This floating sargassum patch is visualized in Fig. (\ref{fig:Sargassum}b) using the Maximum Chlorophyl Index (MCI) \cite{Binding2013}. Due to frequent cloud coverage, such clear snapshots of floating material in the ocean are very rare. \\ \\We use the $ \mathrm{\overline{TRA}} $ to extract the eddy highlighted by the spiralling sargassum patch described in \cite{Beron2015}. We also compute two further single trajectory metrics: the trajectory length diagnostic (Fig. (\ref{fig:Sargassum}c)) and the looping segments derived from the Lagrangian spin (Fig. (\ref{fig:Sargassum}d)). We include a snapshot of the floating sargassum patch which reveals the presence of an elliptic LCS (Fig. (\ref{fig:Sargassum}b)). Similarly to the MCI, the $ \mathrm{\overline{TRA}} $ reveals a Lagrangian eddy centered at $ (67.6^{\circ}W, 37.1^{\circ}N) $. Indeed, this eddy is visible as a distinguished local maximum in the $ \mathrm{\overline{TRA}} $-field surrounded by a dense set of closed and convex contours (zoomed inset of Fig.  (\ref{fig:Sargassum}a)). The white eddy boundary is extracted from the $ \mathrm{\overline{TRA}} $ using the algorithm proposed in Appendix \ref{app: A} with the same parameters presented in section (\ref{sec: QMA}). The extracted eddy boundary underestimates the size of the sargassum patch, but correctly approximates the location (Fig. (\ref{fig:Sargassum}b)). The cyclonic looping exhibited by the two trajectories inside the eddy additionally confirms the presence of an elliptic LCS (Fig. (\ref{fig:Sargassum}d)) but does not provide a specific estimate for the eddy boundary. The trajectory length diagnostic shows a nearby maximum (zoomed inset of Fig. (\ref{fig:Sargassum}c)), which is inconsistent with the generally suggested footprint (local minima) for a coherent eddy in the $ \mathrm{M} $-field (see section (\ref{sec: Mfunction})). \\ \\
\begin{figure}[http]
\centering \includegraphics[scale=0.71]{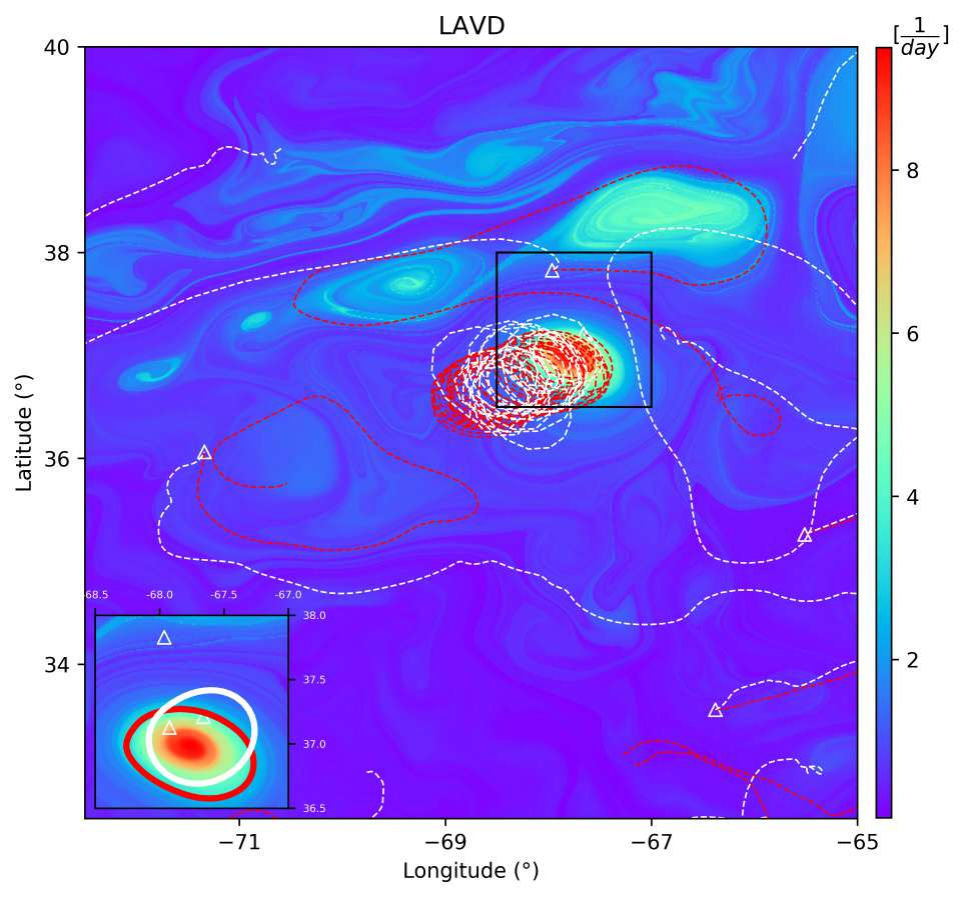} \caption{ \label{fig:Sargassum_LAVD} AVISO-based trajectories (red) and true drifter trajectories (white) superimposed on LAVD-field. The closed white and red curves in the zoomed inset respectively indicate the $ \mathrm{\overline{TRA}} $-and $ \mathrm{LAVD} $-eddy. Triangles denote drifter positions.}
\end{figure}Fig. (\ref{fig:Sargassum_LAVD}) shows the $ \mathrm{LAVD} $-field with AVISO trajectories (red) launched backwards in time from true drifter positions at time $ t =276 \ \mathrm{doy} $. The mesoscale eddy inferred from the sparse drifter distribution (white curve in zoomed inset of Fig. (\ref{fig:Sargassum_LAVD})) is in close agreement with the $ \mathrm{LAVD} $-eddy (red curve in zoomed inset of Fig. (\ref{fig:Sargassum_LAVD})). As expected, drifter and AVISO trajectories show similar looping behaviour inside the identified coherent structure, while they largely differ in other regions.

\section{Conclusion}
Lagrangian eddies (elliptic LCSs) are material objects responsible for the transport of floating particles over large distances in the ocean. They are, by their definition, frame-indifferent and thus can only be reliably deduced from objective feature extraction methods. The local ocean velocity is most accurately observed from float trajectory data, which, however, is inherently non-objective, representing an inconsistency that available eddy detection methods for sparse trajectories do not address. Those methods typically describe eddies by extracting the looping segments of a trajectory, but their definition of looping depends on the frame of the observer. Furthermore, looping segments of a trajectory are most commonly described by these methods in a statistical sense and hence are not geared towards highlighting individual Lagrangian eddy boundaries with high accuracy. \\ \\
In this paper, we have proposed to tackle this inconsistency from a dynamical systems perspective by applying the adiabatically quasi-objective $ \mathrm{\overline{TRA}} $ diagnostic \cite{Haller2021, Haller2021Erratum} to sparse drifter data sets. The $ \mathrm{\overline{TRA}} $ approximates an objective measure of material rotation in frames satisfying specific conditions that generally hold in the ocean. We have found that vortical flow features are related to regions of high local material rotation identified as local maxima in the $ \mathrm{\overline{TRA}} $-field. The $ \mathrm{\overline{TRA}} $ highlights both submesoscale and mesoscale vortices from sparse drifter data, as demonstrated in our two examples. Furthermore, it also succeeds in characterizing the mixing and stirring processes in coastal flow regions and captures the merger of two originally distinct eddies. Contrary to other single trajectory diagnostics, both the $ \mathrm{\overline{TRA}} $ and the $ \mathrm{LAVD} $ are physically related to the local material rotation in the flow. In contrast to the $ \mathrm{LAVD} $, which correctly highlights vortical flow features in sufficiently dense velocity data, the $ \mathrm{\overline{TRA}} $ can be applied to arbitrarily sparse drifter data, given its lack of dependence on nearby drifter trajectories. Hence, it incorporates valuable drifter data into the analysis of oceanic coherent structures in a physically and mathematically justifiable way. This proves to be especially useful in ocean regions where satellite-altimetry data does not unravel the true ocean dynamics. \\ \\ 
Importantly, the looper segments extracted from the Lagrangian spin coincide with the features identified in the $ \mathrm{\overline{TRA}} $. However, a spaghetti plot of looping trajectory segments does not immediately reveal transport barriers and eddy boundaries. Furthermore, potentially valuable information is lost when we discard non-looping trajectory segments based on a manually tuned threshold parameter. The trajectory length diagnostic \cite{Mendoza2010, Mancho2013} also tended to show either minima or maxima near the material eddies highlighted by the local maxima of the $ \mathrm{\overline{TRA}} $-field. This variation in extremum types creates ambiguity in using the trajectory length as a stand-alone indicator for detecting elliptic LCSs from sparse data sets.\\ \\
Apart from the visual inspection of the reconstructed $ \mathrm{\overline{TRA}} $-field, we have additionally presented an algorithm to extract approximate eddy boundaries from sparse drifter data. As vortical flow features are indicated by blobs close to local maxima in the $ \mathrm{\overline{TRA}} $-field, the proposed method resembles a blob detection algorithm. Passing from a continuous scalar diagnostic field to a set of closed curves inevitably requires introducing user-defined parameters. All in all, however, the number of free parameters is comparable to other multi-trajectory Lagrangian eddy detection methods \cite{Haller2016b, Tarshish2018}. This is noteworthy as these algorithms were originally designed assuming knowledge of the underlying velocity field.

\appendix

\section{Eddy boundary extraction algorithm}
\label{app: A}

\begin{algorithm}
\textbf{Input:} Trajectories over the time interval $[t_{0},t_{N}]$. 
\begin{enumerate}
\item Reconstruct $\mathrm{\overline{TRA}_{t_{0}}^{t_{N}}}$-field at time $t$ using linear radial basis interpolation. It is recommended to additionally filter the resulting  $ \mathrm{\overline{TRA}_{t_{0}}^{t_{N}}} $ using a spatial average filter to reduce noise.
\item Find local maxima of $\mathrm{\overline{TRA}_{t_{0}}^{t_{N}}}$ which are above a threshold $ \mathrm{\overline{TRA}_{loc,max}} $.
\item Compute for each closed level set surrounding a local maximum, the averaged $ |\nabla\mathrm{\overline{TRA}_{t_{0}}^{t_{N}}}| $ along the level set.
\item Find closed level set with the highest averaged  $ |\nabla\mathrm{\overline{TRA}_{t_{0}}^{t_{N}}}| $ which additionally
\begin{enumerate}
\item has at least one local maximum of $\mathrm{\overline{TRA}_{t_{0}}^{t_{N}}}$ in its interior.
\item contains at least $ \mathrm{n}_d $-trajectories.
\end{enumerate}
\item Take the convex hull of all the selected closed curves.
\item If two or more convex closed curves intersect, then take the union of these curves.
\item Take the convex hull of the resulting closed curves.
\end{enumerate}
\caption{Extraction of approximate eddy boundaries from $ \mathrm{\overline{TRA}_{t_{0}}^{t_{N}}}$-field}
\textbf{Output:} Approximate eddy boundaries at time
$t$. 
\label{alg: Algorithm1}
\end{algorithm}While various eddy extraction algorithms have been presented in \cite{Williams2015, Hadjighasem2016, Haller2016a, Schlueter2017}, these methods assume a trajectory density that is generally unavailable for drifter trajectories in the ocean. Here, we propose an algorithm that extracts approximate eddy boundaries from the topology of the reconstructed $ \mathrm{\overline{TRA}} $-field. Vortices are then  identified by this algorithm as local maxima of the $ \mathrm{\overline{TRA}} $-field surrounded by a dense set of closed and convex curves characterized by high spatial gradients.\\ \\
Passing from a continuous scalar field to a set of discrete closed curves representing eddy boundaries inevitably requires introducing threshold parameters. There are two main parameters involved in Algorithm \ref{alg: Algorithm1}. The first user-defined quantity aids the identification of the local maxima in the $ \mathrm{\overline{TRA}} $-field. As we identify vortical flow features with regions of high $ \mathrm{\overline{TRA}} $, local maxima below a predefined threshold $ \mathrm{\overline{TRA}_{loc,max}} $ are neglected. Additionally, we also need to specify the minimum number of drifters $ \mathrm{n}_d $ inside an eddy. As elliptic LCSs are often observed via a dense clustering of multiple drifters, this parameter is generally set to be greater than 1. In this work, the parameters are consistently set to $ \mathrm{\overline{TRA}_{loc,max}} = .5 \mathrm{\overline{TRA}_{max}}$, where $ \mathrm{\overline{TRA}_{max}} $ is the global maximum in the $ \mathrm{\overline{TRA}} $-field and $ \mathrm{n_d} = 2 $, as we expect multiple drifters to be part of the same coherent structure. Hence, our eddy extraction algorithm detects elliptic LCSs if at least two trajectories are contained inside an eddy. By construction, eddy boundaries are closed convex curves characterized by sharp gradients . These curves do not necessarily coincide with level sets of the $ \mathrm{\overline{TRA}} $.
\section{Sensitivity analysis with respect to the interpolation method}
\label{app: B}
\begin{figure}[http]
\centering \includegraphics[scale=0.7]{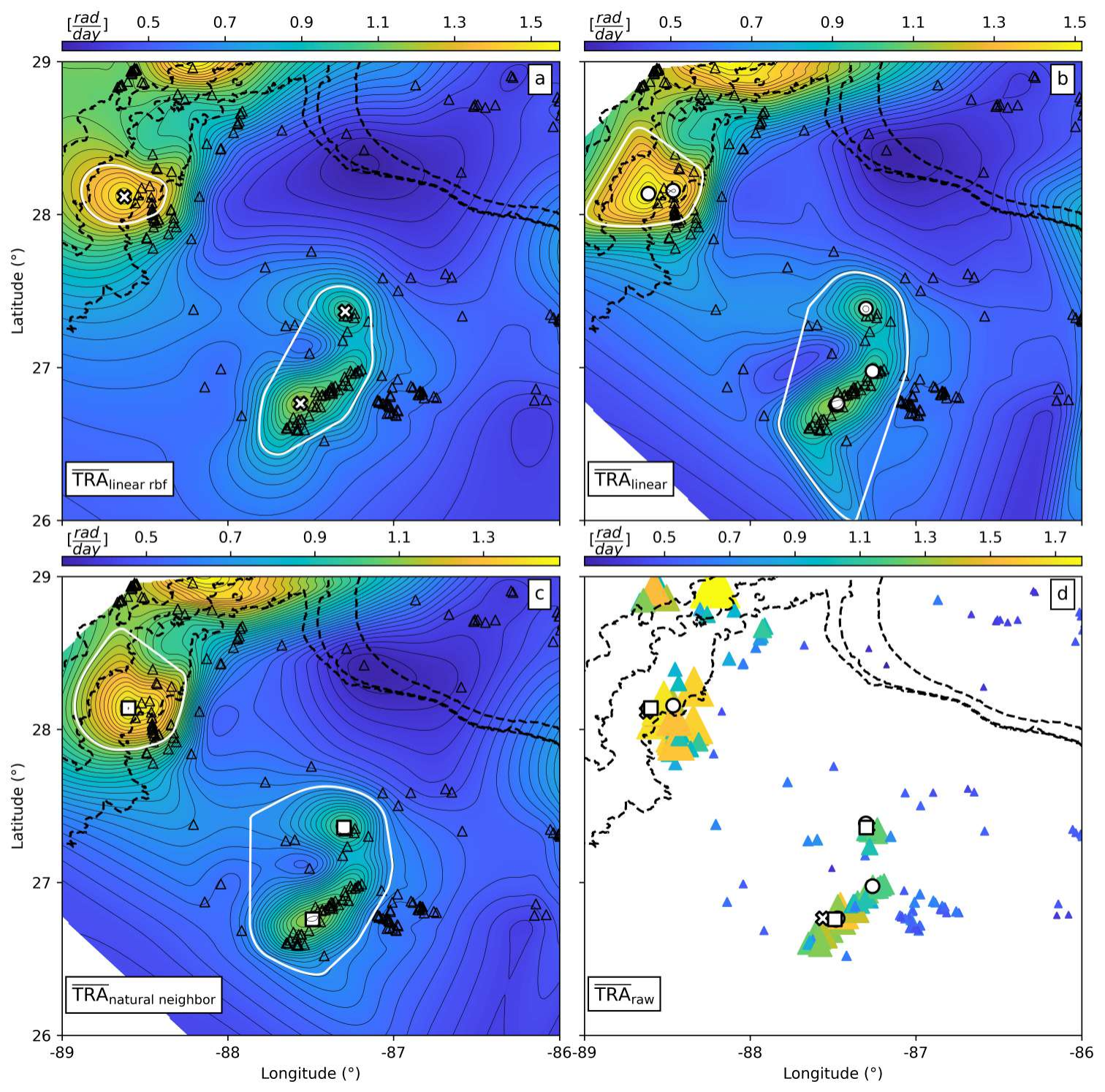} \caption{ \label{fig:TRA_interpolation_GLAD} Comparison of $ \mathrm{\overline{TRA}} $-fields obtained from different interpolation functions with the raw $ \mathrm{\overline{TRA}} $ plot for the GLAD data set at time $ t = 229 \ \mathrm{doy} $. Closed white curves indicate eddies extracted from the reconstructed $ \mathrm{\overline{TRA}} $-field. White crosses, circles and squares denote local $ \mathrm{\overline{TRA}} $ maxima. Triangles indicate the position of the drifters. (a) linear rbf. (b) linear ($ C^{0} $-interpolation) (c) natural neighbor ($ C^{1} $-interpolation) (d) scattered, raw $ \mathrm{\overline{TRA}} $. Color and size of the triangles are related to the $ \mathrm{\overline{TRA}} $-value associated to each drifter.}
\end{figure}

\begin{figure}[http]
\centering \includegraphics[scale=0.7]{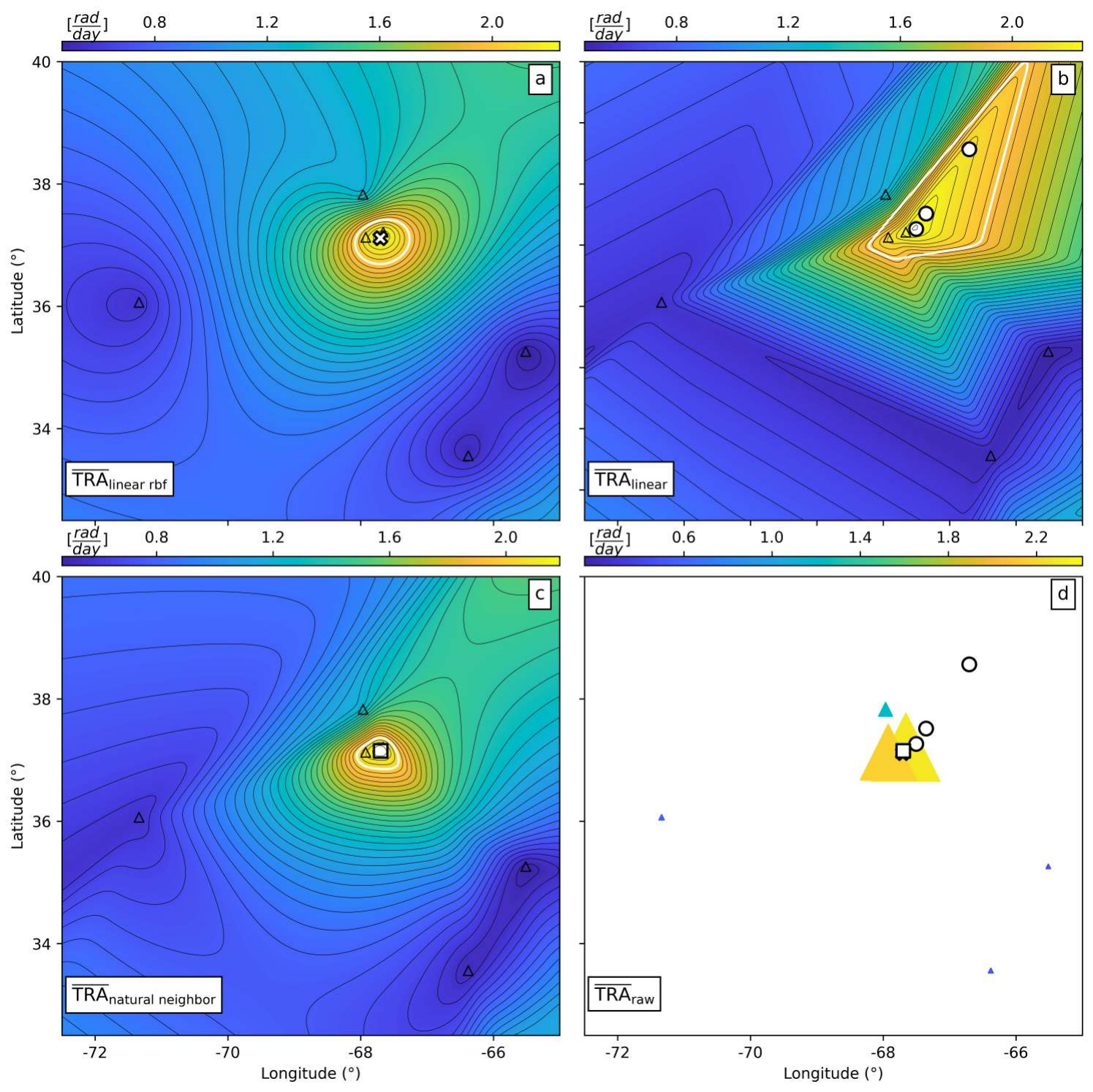} \caption{ \label{fig:TRA_interpolation_GDP} Same as Fig. \ref{fig:TRA_interpolation_GLAD} but for the GDP data set at $ t = 276 \ \mathrm{doy}$.}
\end{figure}

The features of the reconstructed $ \mathrm{\overline{TRA}} $-field clearly depend on the employed interpolation scheme, but we expect the topology of the $ \mathrm{\overline{TRA}} $-field to be robust with respect to the interpolation method in the vicinity of sharp $ \mathrm{\overline{TRA}} $ maxima. Here, we verify the persistence of such local maxima and the extracted $ \mathrm{\overline{TRA}} $-eddies with respect to common interpolants: linear radial basis function (rbf), linear ($C^0$-interpolant) and natural neighbor ($C^1$-interpolant). The reconstructed scalar fields will all be equally pre-and post-processed: inertial oscillations are removed from the drifter trajectories and a spatial averaging filter of size $ (0.25^{\circ} \times 0.25^{\circ}) $ is applied to the reconstructed $ \mathrm{\overline{TRA}} $-field. \\ \\ 
Fig. (\ref{fig:TRA_interpolation_GLAD}) shows the reconstructed $ \mathrm{TRA} $-field using linear rbf, linear and natural neighbour interpolation for the GLAD drifters. For comparison, we also included the raw $ \mathrm{\overline{TRA}} $ plot. Two major eddies are detected independent of the interpolation scheme (see Fig. (\ref{fig:TRA_interpolation_GLAD}a-c)). Both the linear rbf and natural neighbor interpolation reveal three similar local maxima (white crosses/squares in Fig. (\ref{fig:TRA_interpolation_GLAD}a/c)). On average, the white crosses and squares are only separated by $ 1 \mathrm{km} $ ($  \sim 0.01^{\circ}$) and the averaged distance of the local maxima to the closest drifter is around $ 3 \mathrm{km} $ ($  \sim 0.03^{\circ}$) in both cases. \\ \\
The linear interpolant reveals three further local maxima closely located to those suggested by the linear rbf and natural neighbor interpolation scheme (white circles in Fig. (\ref{fig:TRA_interpolation_GLAD}b)). These additional local maxima, however, are not strong enough to lead to the emergence of new eddies. Furthermore, local maxima are located in regions with dense $ \mathrm{\overline{TRA}} $-values (see Fig. (\ref{fig:TRA_interpolation_GLAD}d)). All three interpolation schemes suggest the emergence of two robustly persisting eddies. \\ \\
We perform an analogous sensitivity analysis on the reconstructed $ \mathrm{\overline{TRA}} $-field for the GDP drifters at time $ t = 276 \ \mathrm{doy} $ (see Fig. (\ref{fig:TRA_interpolation_GDP})). Local maxima of the linear rbf and natural neighbour interpolation are in close agreement with each other, whereas the linear interpolant introduces two further local maxima. Again, these additional local maxima are weak and do not lead to the detection of additional eddies. In summary, the eddy boundary extraction algorithm consistently reveals a mesoscale eddy independent of the interpolation scheme. \\ \\  
Overall, we find that the eddy location does not strongly depend on the interpolation scheme  at least in the two data sets considered in this paper. However, the exact shape and size of the eddies vary as a function of the interpolation method. In very sparse data sets, such as the GDP-dataset, the identified eddy area may vary substantially under changes in the interpolation method. The location of the largest local $ \mathrm{\overline{TRA}} $ maximum signaling the presence of an eddy, however, is fairly robust with respect to the interpolation method. Due to its inherent radial symmetry, linear rbf interpolation tends to favour elliptic structures. In contrast, linear and natural neighbour interpolation lead to sharper and non-smooth eddy boundaries.
\newpage 
\textbf{Data Availability} \\ \\
The AVISO geostrophic current velocity product used in this study "Global Ocean Gridded L4 Sea Surface
Heights and Derived Variables Reprocessed" is freely available and is hosted by the Copernicus Marine Environment
Monitoring Service \cite{CMEMS}. 
The GDP drifter data is openly available in the \href{http://www.aoml.noaa.gov/phod/dac/}{NOAA Global Drifter Program dataset} \cite{Lumpkin2019}. The GLAD drifter data is distributed by the \href{https://data.gulfresearchinitiative.org/data/R1.x134.073:0004}{Gulf of Mexico Research Initiative Information and Data Cooperative (GRIIDC)} \cite{Ozgokmen2013}.
The attracting LCS and LAVD computations have been performed with the software package \href{https://github.com/haller-group/TBarrier}{TBarrier}.
Jupyter notebooks (together with drifter data) implementing the methods used here are available under \href{https://github.com/EncinasBartos/QuasiObjectiveEddyVisualizationFromSparseDrifterData}{Github/EncinasBartos}. These notebooks can readily be applied to any sparse trajectory data set. \\ \\
\newpage
\textbf{Acknowledgements} \\ \\
The authors acknowledge financial support from Priority Program SPP 1881 (Turbulent Superstructures) of the German National Science Foundation (DFG).


\end{document}